\documentclass[11pt,a4paper]{article}
\usepackage{jinstpub}

\usepackage{xspace}
\usepackage{upgreek}
\usepackage{makecell}

\begin{document}


\newcommand{\neqcm}{\ensuremath{\,\mathrm{n_{eq}/cm^2}}\xspace}
\newcommand{\cm}{\ensuremath{\,\mathrm{cm}\xspace}}
\newcommand{\micron}{\ensuremath{\,\mathrm{\mu m}}\xspace} 
\newcommand{\cmsq}{\ensuremath{\,\mathrm{cm^2}}\xspace}
\newcommand{\msq}{\ensuremath{\,\mathrm{m^2}}\xspace}
\newcommand{\Ialpha}{\ensuremath{\alpha}\xspace}
\newcommand{\degreeC}{\ensuremath{\,\mathrm{^\circ C}}\xspace}
\newcommand{\minute}{\ensuremath{\,\mathrm{min}}\xspace}
\newcommand{\minutes}{\ensuremath{\,\mathrm{min}}\xspace}
\newcommand{\kiloHertz}{\ensuremath{\,\mathrm{kHz}}\xspace}
\newcommand{\Volt}{\ensuremath{\,\mathrm{V}}\xspace}
\newcommand{\Neff}{\ensuremath{{\mathrm{N_\text{eff}}}}\xspace}
\newcommand{\Neffnorad}{\ensuremath{{\mathrm{N_\text{eff,0}}}}\xspace}
\newcommand{\DeltaNeff}{\ensuremath{{\Delta N_\text{eff}}}\xspace}
\newcommand{\Udep}{\ensuremath{U_\text{dep}}\xspace}

\newcommand{\MeV}{\ensuremath{\,\mathrm{MeV}}\xspace}

\newcommand{\dCell}[2]{\vtop{\hbox{ \strut #1}\hbox{\strut #2}}}

\title{Isothermal annealing of radiation defects in silicon bulk material of diodes from 8" silicon wafers}

\author[a,d,1]{Jan Kieseler \note{Corresponding author.}}
\author[a,b]{Pedro Gon\c{c}alo Dias de Almeida}
\author[a,c,d]{Oliwia Ka\l{}uzi\'{n}ska}
\author[a,f]{Marie Christin M\"uhlnikel}
\author[a]{Leena Diehl}
\author[a]{Eva Sicking}
\author[a,e]{Philipp Zehetner}

\affiliation[a]{CERN, Geneva, Switzerland} 
\affiliation[b]{CSIC, Santander, Spain}
\affiliation[c]{Wroc\l{}aw University of Science and Technology, Wroc\l{}aw, Poland}
\affiliation[d]{Karlsruhe Institute of Technology, Karlsruhe, Germany}
\affiliation[e]{Ludwig-Maximilians-Universit\"at München, Munich, Germany}
\affiliation[f]{University of Heidelberg, Germany}

\emailAdd{jan.kieseler@cern.ch}
\date{\today}


\abstract{
  The high luminosity upgrade of the LHC will provide unique physics opportunities, such as the observation of rare processes and precision measurements. However, the accompanying harsh radiation environment will also pose unprecedented challenged to the detector performance and hardware. In this paper, we study the radiation induced damage and its macroscopic isothermal annealing behaviour of the bulk material from new 8" silicon wafers using diode test structures. The sensor properties are determined through measurements of the diode capacitance and leakage current for three thicknesses, two material types, and neutron fluences from $6.5\cdot 10^{14}$ to $1 \cdot 10^{16}\,\mathrm{n_{eq}/cm^2}$. 
}

\maketitle

\section{Introduction}

The high-luminosity upgrade of the CERN LHC (HL-LHC)~\cite{Evans_2008,Apollinari:2284929} will allow the experiments located at the interaction points to accumulate an unprecedented amount of data which will open new opportunities for physics studies, as well as provide a key tool for high precision measurements. The required high instantaneous luminosity is accompanied by up to 200 simultaneous interactions (pileup) per bunch crossing, which is almost an order of magnitude larger than the current running conditions. To maintain or exceed the physics performance of the LHC Run 1 to 3, it is crucial to upgrade the detectors and adapt new reconstruction techniques. An important aspect of these upgrades are radiation tolerance considerations adapted to the harsh radiation environment. The radiation levels will increase by about a factor of 10~\cite{Apollinari:2284929}.
In particular in the forward region at pseudorapidities above 1.5, this calls for a detector design that can operate under these conditions and offers the possibility to use fine-grained particle flow and calorimetry algorithms, such that the primary interaction particles can be separated from the pileup interactions using information from all detector subsystems~\cite{CMSPFPaper,ATLASPF,Rovere:2020rqi,qasim2022}.

The high-granularity calorimeter (HGCAL), planned to replace the current CMS~\cite{CMS_det_paper} endcap calorimeters for the HL-LHC upgrade~\cite{HGCAL-TDR}, is a sampling calorimeter consisting of 47 sensor and absorber layers corresponding to a total of nearly 10 hadronic interaction lengths. It covers the pseudorapidity range of 1.5 to 3.0, and is placed 320\cm\ from the beam spot in z. Here, the coordinate system is defined with the beam spot at its centre. The z axis follows the beam direction.
The calorimeter is roughly divided into three parts in depth. The first part comprises 26 very fine sampling layers and corresponds to 27 electromagnetic radiation lenghts and provides granularities down to cell areas of only 0.5\cmsq. It is followed by 11 fine sampling layers and the remaining 10 layers with more absorber material. 
The entire first 33 layers, placed at $|z| \lesssim 410\cm$, as well as the other parts closer to the beam line than approximately 90\cm\ are subject to higher radiation levels above $10^{15}$ 1\MeV equivalent neutron fluence per $\mathrm{cm^2}$ (\neqcm) and are equipped with silicon sensors as active material. Further from the interaction point, scintillator tiles are used to detect the deposited energy with a fixed size in pseudorapidity and polar angle $\Phi$, defined in the x-y plane, of $0.022 \times 0.022$. The silicon sensors will be subject to radiation damage under these harsh conditions with up to $1 \cdot 10^{16} \neqcm$. To reduce the damage to the bulk material, silicon diodes with different thicknesses of 120, 200, or 300\micron are placed such that the thickness decreases with increasing expected fluence. The 120\micron sensors are made of p-type epitaxial (EPI) material, while the other thicknesses use the p-type float zone (FZ) base material. This choice is motivated by manufacturing constraints.
The p-type sensors have been proven to be favourable in high radiation environements as they do not undergo type inversion, appear to be more resistant to non-Gaussian noise due to junction breakdown, and provide fast readout at n-type electrodes~\cite{Peltola_2015,CASSE2002465,CASSE200646}.



To cover the large area of silicon in the HGCAL (more than 600\msq) in a cost-efficient manner, ingots with 8" diameter are employed as base material, making HGCAL the first High-Energy Physics particle detector project using silicon sensors produced in an 8" process. The sensors are produced on corresponding 8" wafers by Hamamatsu Photonics~\cite{hamamatsu}. The HGCAL sensors have hexagonal shape, and can therefore exploit the area of each wafer more optimally than square sensors. The 8" wafer base material has different characteristics than the 6" wafer material used for the CMS tracker and its upgrades. Therefore, not all conclusions drawn from a variety of previous measurements of the effect of radiation damage and annealing behaviour based on the 6" wafer material might apply, requiring a novel qualification of the radiation hardness of the 8" material. A parameterisation of detector behaviour with radiation is needed to predict the behaviour of the 8" sensors during the lifetime of the HGCAL and to establish operating conditions that guarantee successful operation during the full lifetime of HL-LHC.

In this paper, we document for the first time the macroscopic effects of radiation damage, and subsequent isothermal annealing of the introduced defects in the base crystal of these 8" wafers, and investigate the annealing behaviour of the bulk material and its parameterisation using dedicated diode test structures. For this purpose, we measure capacitance and leakage current as a function of the bias voltage for a set of annealing steps and different fluences from $10^{15}$ to $10^{16}\neqcm$ for all thicknesses. Furthermore, we study the radiation induced leakage current and the behaviour of the depletion voltage as a function of annealing time according to the Hamburg model~\cite{Lindstrom:1999mw,Moll:1999kv}.

The paper is organised as follows: in Section~\ref{seq:samples}, we describe the samples, their thicknesses and materials in detail. The measurements of leakage current and capacitance are described in Section~\ref{seq:measurements}. The interpretation of the data in terms of the Hamburg model and the current related damage rate, \Ialpha, can be found in Section~\ref{seq:interpretation}. Finally, we conclude the paper in Section~\ref{seq:conclusions}.

\section{Samples}
\label{seq:samples}

The diode test structures are located at the upper and lower corners of the wafers that are cut off to retrieve the full hexagonal-shaped sensor. Here, we measure the structures in the upper right (UR) and upper left (UL) corners. Further, we distinguish individual wafers by a 4-digit production number. The diodes are cut from the wafers before they are irradiated with neutrons at the 
\textit{Training, Research, Isotopes, General Atomics (TRIGA)} reactor located at the Jo\v{z}ef Stefan Institute, Ljubljana, Slovenia~\cite{dimic_reactor_1978}. 

\begin{figure}[htbp]
\begin{center}
\includegraphics[width=0.58\textwidth]{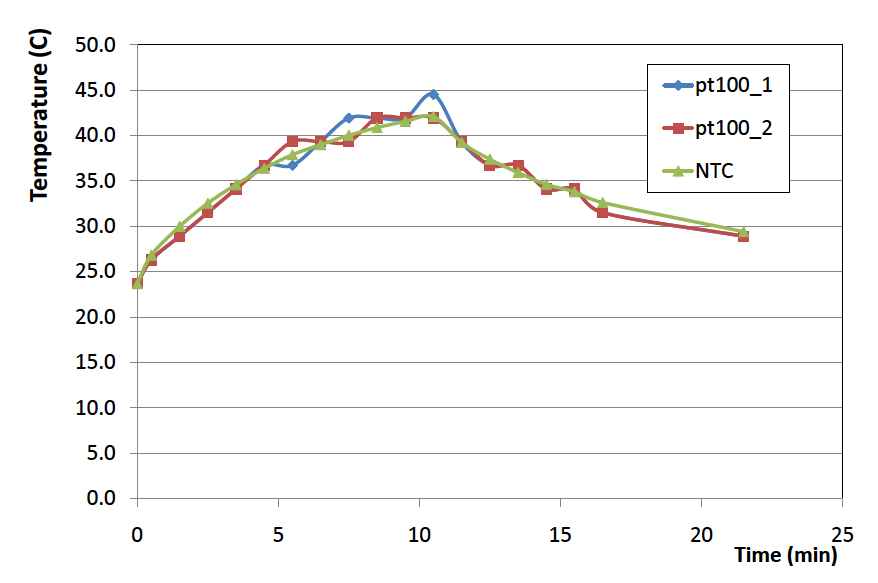}
\caption{Temperature profile measured by different temperature probes for an irradiation time of 10.8\minute in the TRIGA reactor measured by different, independent sensors, taken from Ref.~\cite{Cindro:2019cxd}. The irradiation time corresponds to a fluence of $10^{15} \neqcm$.}
\label{fig:temp_prof}
\end{center}
\end{figure}

During radiation, the samples are brought to temperatures where the annealing effects are not negligible and results need to be corrected for this effect when extracting the annealing behaviour. Therefore, the effective annealing during irradiation and subsequent cooling is calculated from the temperature profiles in the container holding the samples within the reactor~\cite{Cindro:2019cxd}. While Ref.~\cite{Cindro:2019cxd} only provides a temperature profile for a specific fluence, shown in Figure~\ref{fig:temp_prof}, the profiles for other fluences can be inferred from the total irradiation time, a maximum temperature between 45\degreeC and 55\degreeC in the container, and a room temperature of 25\degreeC. We assume that the temperature rise and decrease can be expressed by an exponential function, each with different time constants. We fit these time constants to the provided profile for one fluence and extrapolate to the other fluence points based on the fitted functional forms. Furthermore, we consider a linear rise in time of the total fluence the material is subjected to during exposure. 

The resulting annealing offset times are converted to equivalent times at 60\degreeC based on the parametrisation derived in Ref.~\cite{Moll:1999kv} and are listed in Table~\ref{tab:samples} together with other sample parameters. We derive an uncertainty in the annealing time offsets by considering the full effect of the uncertainty in the reactor temperature, which has a larger effect for longer annealing times.

\begin{table}[]
    \centering
    \caption{Overview of samples considered for the studies described in this paper, where FZ stands for float zone and EPI for epitaxial. The labels ``U'' (upper right), ``UL'' (upper left), and ``LL'' (lower left) refer to the placement of the test structures on the wafer. The uncertainty on the fluence is 10\% for all fluences. The corresponding annealing offsets are listed as ``Offset''.}
    \label{tab:samples}
    \renewcommand{\arraystretch}{1.5}
    \begin{tabular}{c|c|c|c|c|c|c|}
         Number  & Position & Material & \dCell{Thickness}{[\micron]} & \dCell{Irradiation time}{[\minute]} & \dCell{Fluence}{[$10^{15} \neqcm$]} & \dCell{ Offset}{[\minute]} \\
        \hline
        1002          & UL+UR+LL    & FZ       & 300       & 7.4          & 0.65                     & 1.0 $\pm^{0.8}_{0.4}$ \\
        1003          & UL+UR       & FZ       & 300       & 10.8         & 1.00                     & 1.6 $\pm^{1.3}_{0.7}$ \\
        1102          & UL+UR+LL    & FZ       & 300       & 16.2         & 1.50                     & 2.4 $\pm^{2.1}_{1.1}$ \\
        2002          & UL+UR+LL    & FZ       & 200       & 10.8         & 1.00                     & 1.6 $\pm^{1.3}_{0.7}$ \\
        2003          & UL+UR       & FZ       & 200       & 16.2         & 1.50                     & 2.4 $\pm^{2.1}_{1.1}$ \\
        2102          & UL+UR+LL    & FZ       & 200       & 27.5         & 2.50                     & 4.0 $\pm^{3.8}_{1.9}$ \\
        3008          & UL+LL       & EPI      & 120       & 16.2         & 1.50                     & 2.4 $\pm^{2.1}_{1.1}$ \\
        3007          & UL          & EPI      & 120       & 27.5         & 2.50                     & 4.0 $\pm^{3.8}_{1.9}$ \\
        3003          & UL+LL       & EPI      & 120       & 108          & 10.00                    & 14.2 $\pm^{14.2}_{7.2}$ \\
    \end{tabular}
\end{table}

\begin{figure}
    \centering
    \includegraphics[width=0.5\textwidth]{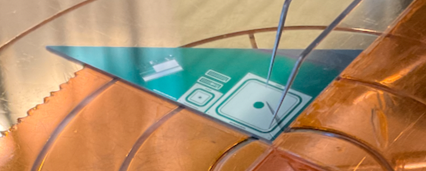}
    \caption{Test structures on a temperature controlled vacuum chuck. The largest diode is connected to the instruments through needles. The second needle establishes contact between the guard ring and ground. The bias voltage is supplied through the chuck.}
    \label{fig:diode}
\end{figure}

All samples have been produced with the target to match the properties of the 6" wafers used in the CMS outer tracker, where possible, also aiming at -2\Volt flat band voltage. Further details are part of the production process and are not disclosed by the manufacturer. The test structures silicon shreds contain more than one diode as shown in Figure~\ref{fig:diode}. For the measurements, we select the largest one with an area of $0.2595\cm^2$ surrounded by a guard ring. 

\section{Measurements}
\label{seq:measurements}

For each sample, we measure the leakage current (IV) and the capacitance (CV) as a function of reverse diode bias voltage for different annealing steps. All measurements are performed at -20\degreeC with diodes directly placed on a vacuum chuck in a Faraday cage, flooded with dry air. 
The capacitance is determined using a frequency of 10\kiloHertz for the UL and UR samples. The LL samples are only used for a dedicated frequency study described in Section~\ref{seq:interpretation}. The measurement is performed with an amplitude of 0.5\Volt, which provides stable results for the particular measurement setup. The leakage current and the capacitance are measured through two independent sets of instruments. For each of the measurements the guard ring is connected to ground. The bias voltage is varied from 0\Volt to up to 1k\Volt, where possible. For some sample, the voltage is restricted to below 900\Volt to avoid too large currents through the diode.

\begin{figure}[htbp]
    \centering
    \includegraphics[width=0.48\textwidth]{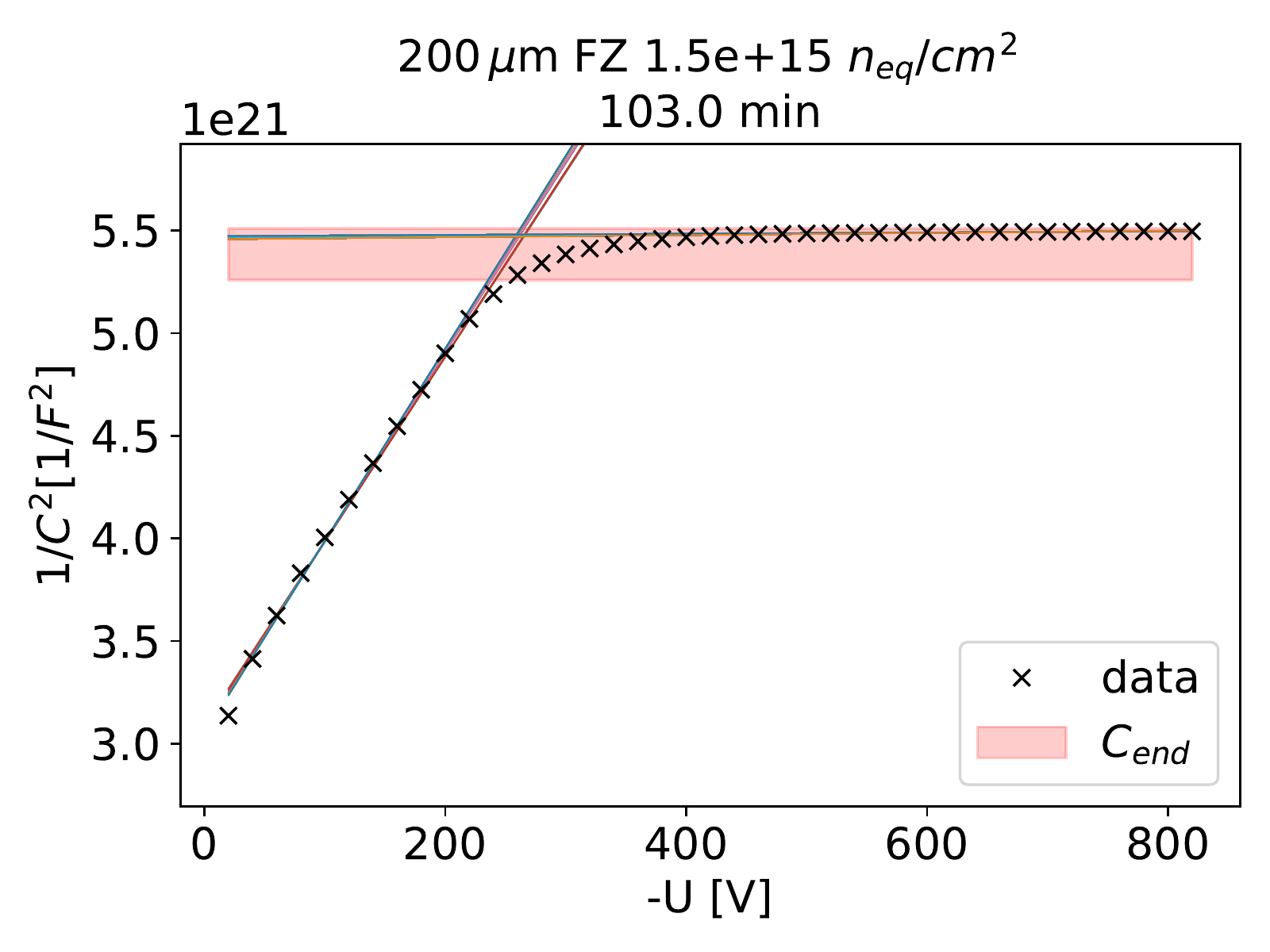}
    \includegraphics[width=0.48\textwidth]{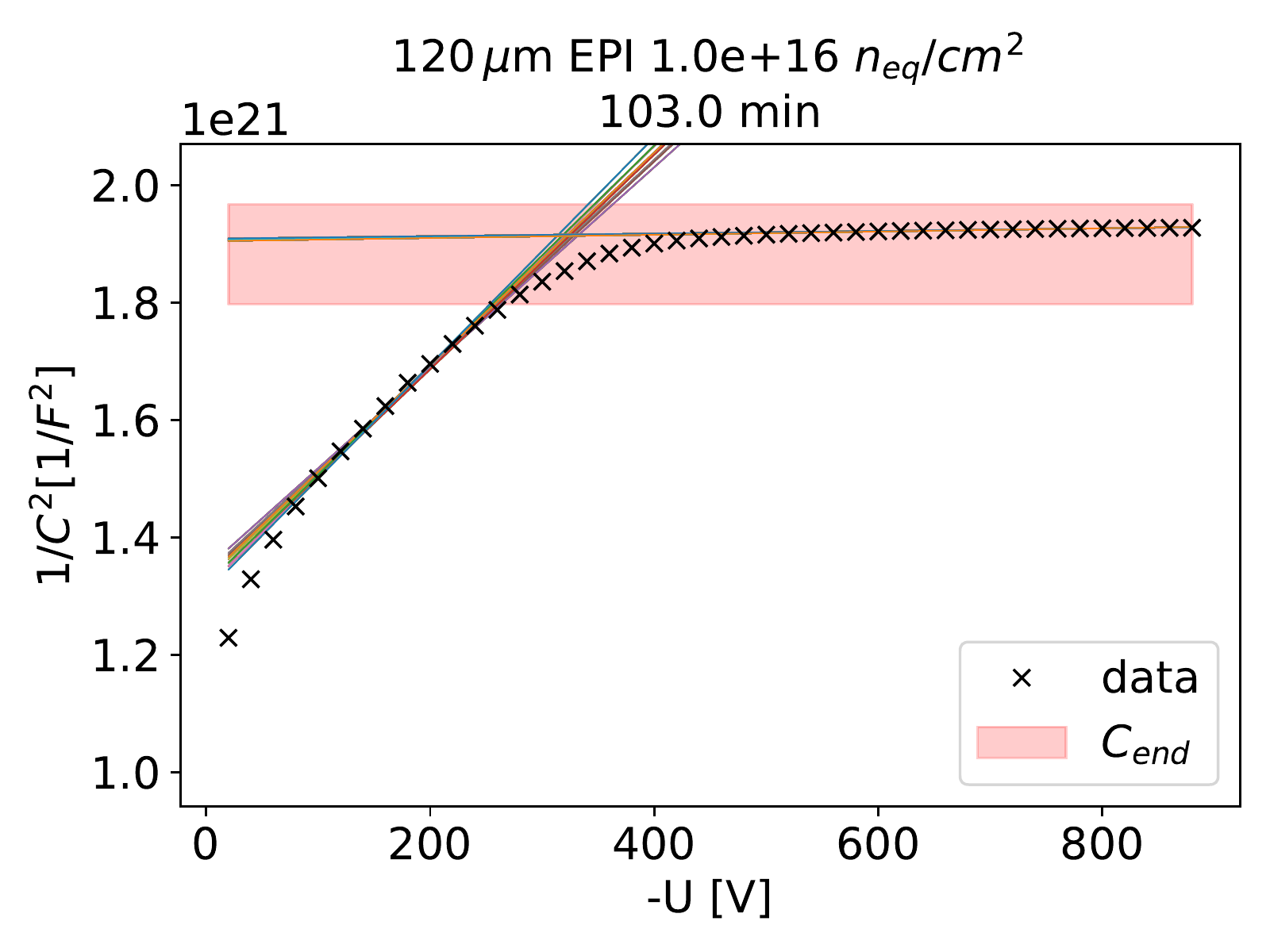}
    \caption{Fit of the depletion voltage from the dependence of the capacitance on the bias voltage. The markers indicate measured points, while the lines show the different fitted curves corresponding to variations of the fit ranges. Left: a sample at low fluence, right: the highest fluence sample.}
    \label{fig:CV_examples}
\end{figure}

The capacitance and conductivity are measured by the LCR meter in parallel mode, but the capacitance is converted to serial mode, $C_S$ as it provides a better approximation of the measured circuit for this frequency choice and for irradiated sensors.
The depletion voltage is extracted from the CV curve by fitting two linear functions to the $1/C_S^2$ versus bias voltage curve; one is fitted to the rising edge, and one to the plateau region, as shown in Figure~\ref{fig:CV_examples}. A 10\% uncertainty is assigned to the chosen fit ranges and alternative fits are performed for each combination of fit ranges. For the highest fluence sample, the uncertainty on the chosen fit range is increased to 20\% as the rising edge deviates from a linear function significantly. This variation covers the possible interpretations of the rising edge as a linear function. As shown in Figure~\ref{fig:CV_examples}, this leads to a significant increase in the uncertainty compared to lower fluence samples.

In case the plateau cannot be reached within the considered bias voltage range, a fixed end capacitance is assumed. This applies only to some 300 and 200\micron samples. The assumption of a fixed end capacitance that does not depend on the annealing time is verified using those samples where a plateau can be reached. To determine the end capacitance for each thickness, only samples with depletion voltages below and absolute value of 600\Volt are used, such that there are sufficient data points in the plateau region. Further, a subset of measurements is selected where the open correction was derived exactly before the measurement was performed.
The uncertainty on the end capacitance is derived from the one sigma spread of the measured points, and is extended if necessary to cover the difference to the measured unirradiated capacitance. The results are compared to the calculated capacitance in Table~\ref{tab:caps}. The measured capacitances are above the calculated capacitance in all cases, which might indicate a slightly smaller effective thickness.

\begin{table}[h]
    \centering
    \caption{Capacitance (expressed as $1/C_S^2$) in the fully depleted regime for each sensor thickness.}
    \label{tab:caps}
    \begin{tabular}{c|c|c|c}
         Nominal thickness  & Calculated [$10^{21}/F^2$] & Unirradiated [$10^{21}/F^2$] & Measured [$10^{21}/F^2$] \\
         \hline
         120\micron & 2.00 & 1.80 & $1.88 \pm 0.08$ \\
         200\micron & 5.56 & 5.51 & $5.38 \pm 0.13$ \\
         300\micron & 12.5 & 11.8 & $11.5 \pm 0.3$ \\
        \hline
    \end{tabular}
\end{table}

\begin{figure}[htbp]
    \centering
    \includegraphics[width=0.48\textwidth]{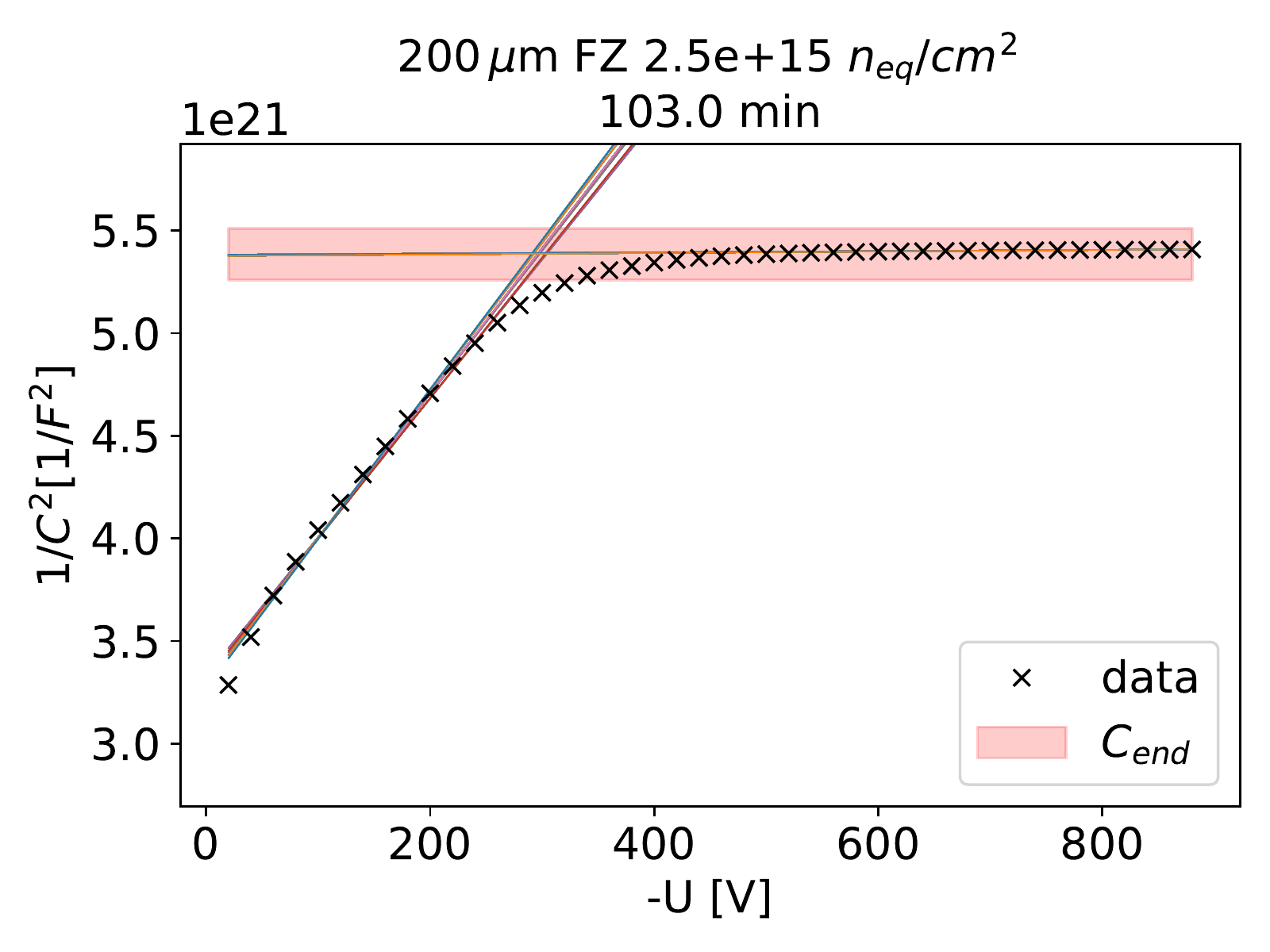}
    \includegraphics[width=0.48\textwidth]{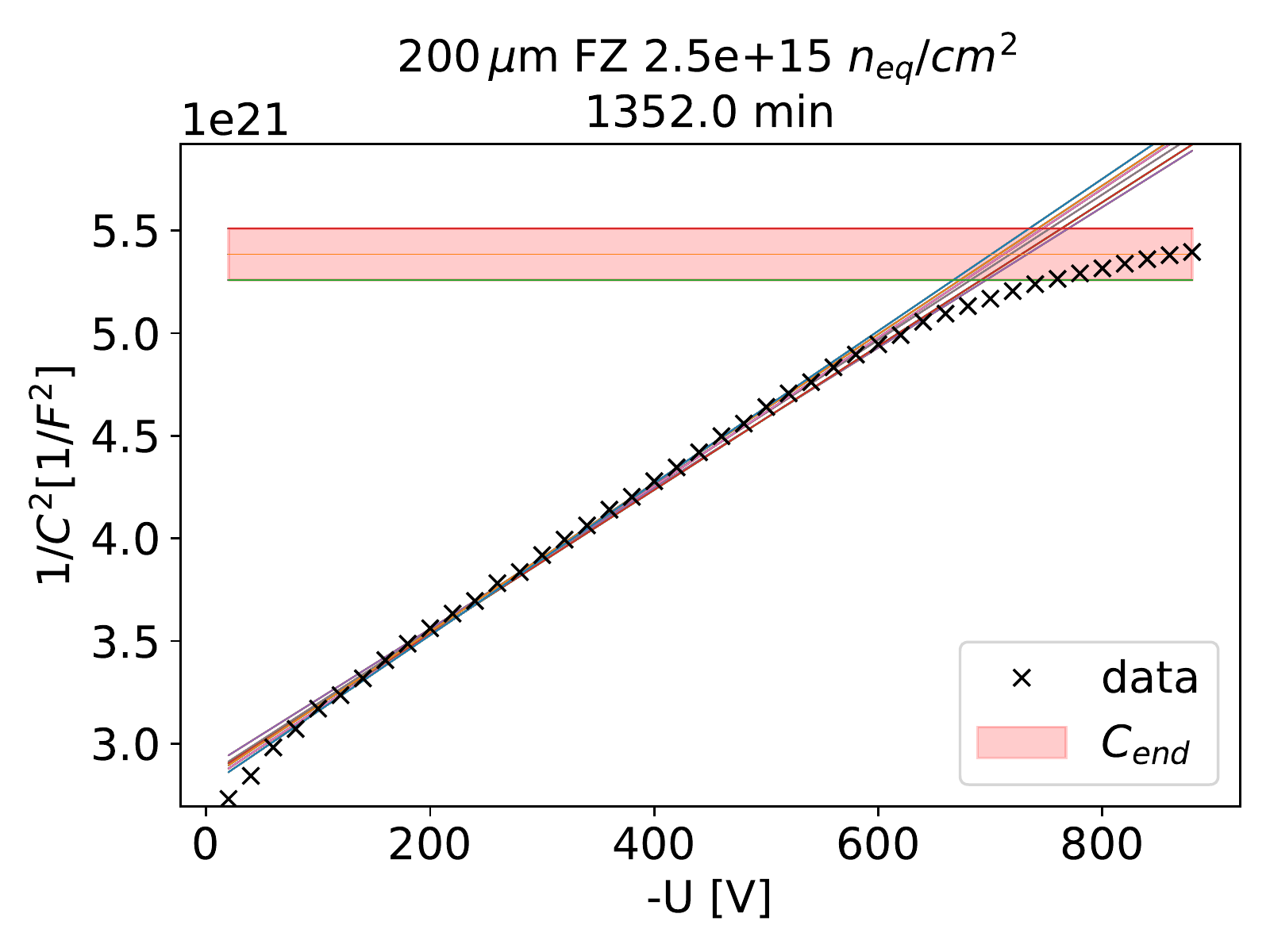}
    \includegraphics[width=0.48\textwidth]{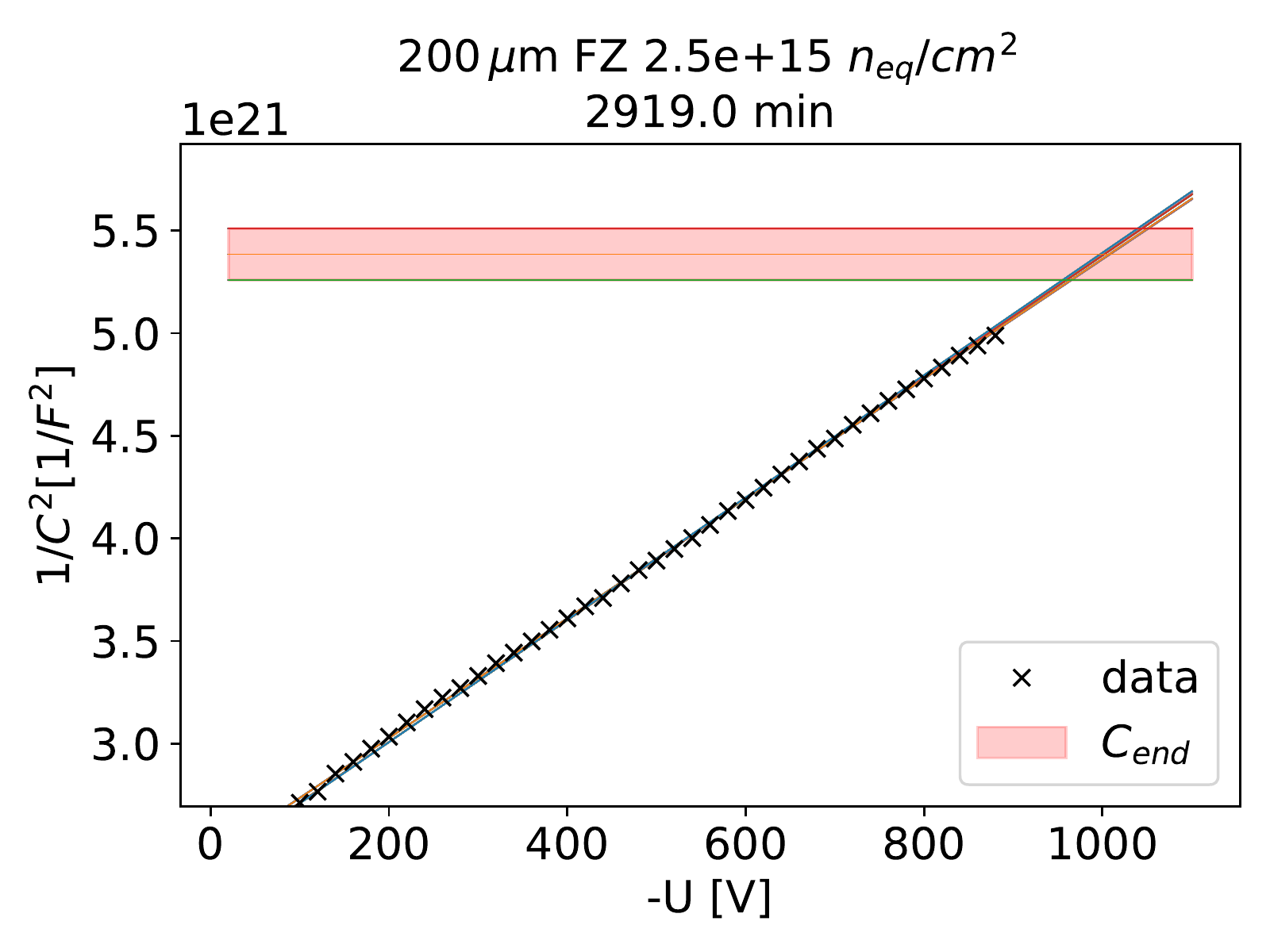}
    \caption{Fit of the depletion voltage from the dependence of the capacitance on the bias voltage. The markers indicate measured points, while the lines show the different fitted curves corresponding to variations of the fit ranges. Top left: plateau region within the measured range; top right: plateau region slightly outside the measured range; bottom: extrapolation beyond the measured range.}
    \label{fig:CV_plateau}
\end{figure}

An example of the depletion voltage fit is shown in Figure~\ref{fig:CV_plateau}, where the plateau region is either determined from the measured data points or from the end capacitance.
 The total uncertainty of the extracted depletion voltage is determined as the envelope of all possible intersection points, including variations of the plateau and the rising edge parameterisation.

The leakage current is measured with a pico-ampere meter and quantifies the current that flows between the silicon pad and the backside while the guard ring is connected to ground.
Even though the current is measured for a fine-binned set of bias voltages, its remaining non-continuities when expressed as a function of the bias voltage are mitigated by applying a Savitzky-Golay filter~\cite{SavGol} with a window length of 5 points and a cubic polynomial interpolation to the data points.

Leakage current and capacitance are obtained for a set of target annealing steps summarised in Table~\ref{tab:ann}. The first steps were performed at 60\degreeC, and the last at 80\degreeC. 
The  measurements were organised in two campaigns. For the first campaign, the 200 and 300\micron UL samples were annealed and measured. Here, the second and third annealing step have been omitted. In the second campaign the 120\micron thick UL samples and the remaining UR samples were studied, including all steps at lower annealing times, but omitting some of the longer annealing times for the diodes with 200 and 300\micron thickness as the measurements proved to be consistent with the ones taken in the first campaign.

The annealing is performed in a temperature controlled oven, by placing the diodes on a pre-heated copper plate. Since the temperature is not constant, in particular due to placing the diodes in, and removing them from the oven, the temperature is recorded every second through a \textit{PT1000} temperature sensor (with 0.1\degreeC intrinsic uncertainty) attached to a silicon piece of similar size as the diodes that is placed in the oven next to them. The uncertainty in the time measurement is negligible. For the temperature measurement, we assign 0.2\degreeC uncertainty to this procedure, which translates to about 1\minute uncertainty when calculating the equivalent annealing time at 60\degreeC. Deviations from the target annealing times are found to be small, but non-negligible. Therefore, the actual annealing times are propagated to the results, and deviations from the target temperature of 60\degreeC are corrected for using the parametrisation for the annealing-time dependent leakage current from Ref.~\cite{Moll:1999kv}. The same parametrisation is used when converting the annealing steps at 80\degreeC to 60\degreeC, leading to an acceleration factor of about 14. We apply the same conversion factor for the leakage current measurements and for the studies of the annealing behaviour of the depletion voltage and all derived quantities, even though the conversion factors differ by up to 5\%~\cite{Moll:1999kv}. This difference is considered an additional uncertainty on the annealing time and has negligible impact on the final results. We consider it part of future studies to confirm the exact time constants for this particular wafer material.

\begin{table}[]
    \centering
    \caption{Target annealing steps. The annealing steps at 80\degreeC are converted to the equivalent time at 60\degreeC using the parameterisation from Ref.~\cite{Moll:1999kv} and is based on the leakage current behaviour.}
    \label{tab:ann}
    \begin{tabular}{c|c|c}
      Total time at 60\degreeC [\minute]   &  Annealing step [\minute] & Temperature [\degreeC]\\
      \hline
      10 & 10 & 60 \\
      30 & 20 & 60 \\
      70 & 40 & 60 \\
      150 & 80 & 60 \\
      250 & 100 & 60 \\
      390 & 140 & 60 \\
      \hline
      660 & 25 & 80 \\
      1450  & 60 & 80 \\
      3000 & 120 & 80 \\
    \end{tabular}
\end{table}

\begin{figure}[htbp]
    \centering
    \includegraphics[width=0.48\textwidth]{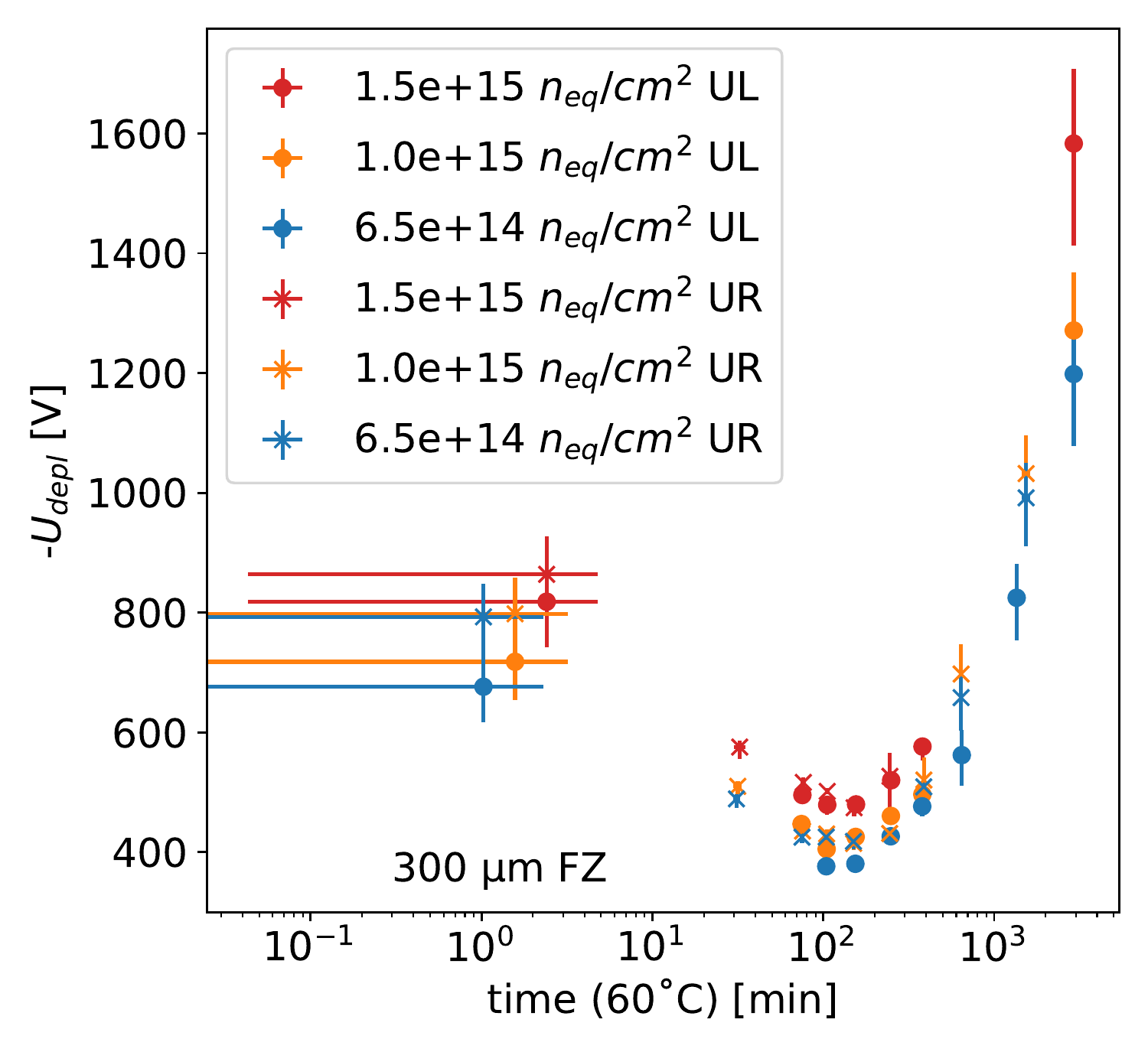}
    \includegraphics[width=0.48\textwidth]{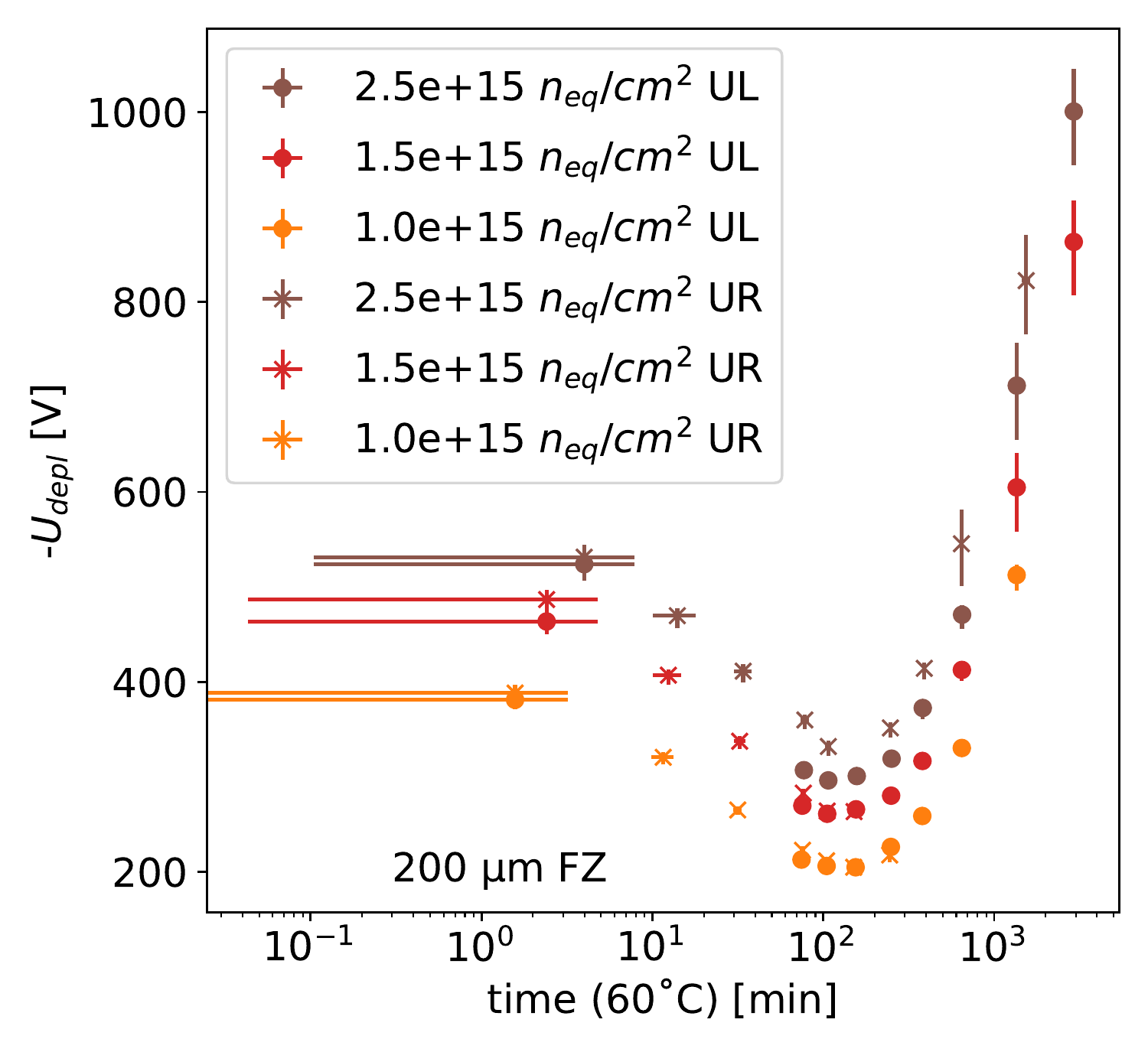}
    \includegraphics[width=0.48\textwidth]{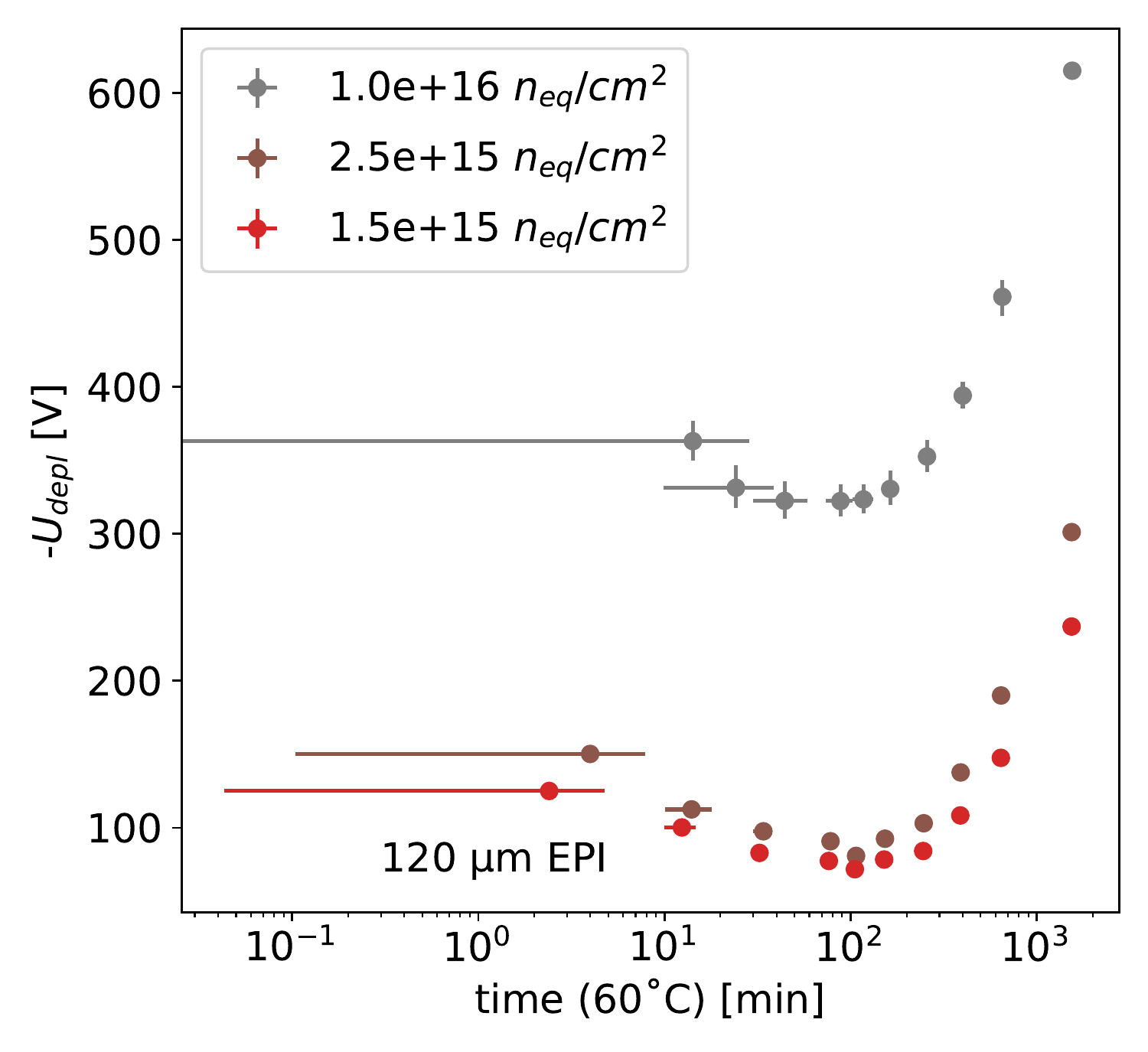}
    \caption{Depletion voltage for the FZ diodes with 300 (upper left), 200 (upper right) \micron thickness, and the EPI diodes (bottom) with 120\micron thickness as a function of the equivalent annealing time at 60\degreeC.}
    \label{fig:deplvolt}
\end{figure}

The depletion voltages extracted from the data are shown in Figure~\ref{fig:deplvolt} for different sensor thicknesses and fluences. The float zone material (200 and 300\micron thickness) shows a qualitatively very consistent behaviour. There are small discrepancies between the measurements from the UL and UR diodes, even for the same nominal fluences. However, the samples were not irradiated together, and a spread in fluence of up to 10\% is expected for the TRIGA reactor. The minimum depletion voltage is reached around 130\minute equivalent annealing time at 60\degreeC. The epitaxial material seems to have a minimum at slightly lower annealing times, but the measurements are otherwise consistent with what is observed for the 200 and 300\micron thick float-zone diodes. 

\begin{figure}[ptbh]
    \centering
    \includegraphics[width=0.32\textwidth]{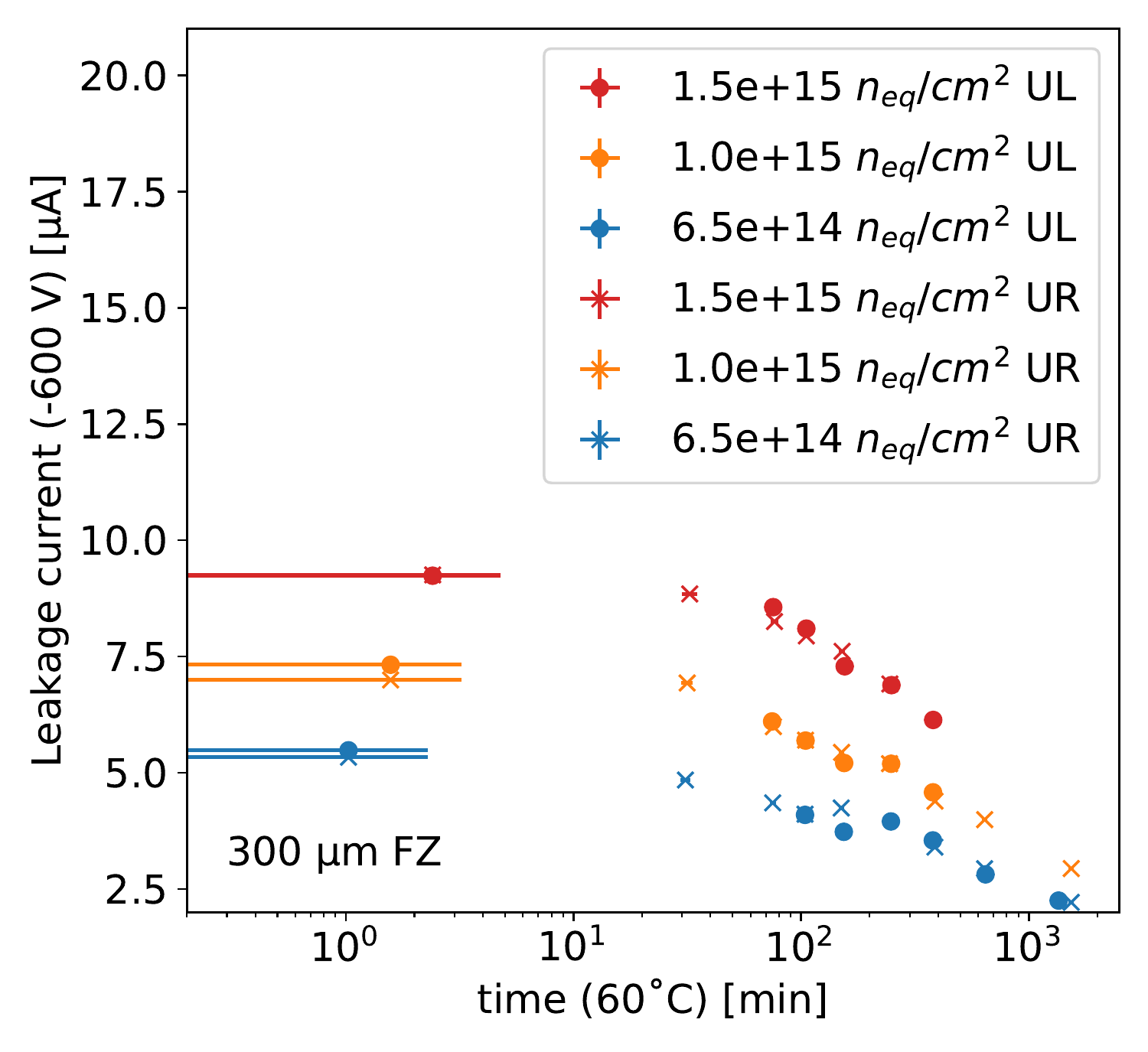}
    \includegraphics[width=0.32\textwidth]{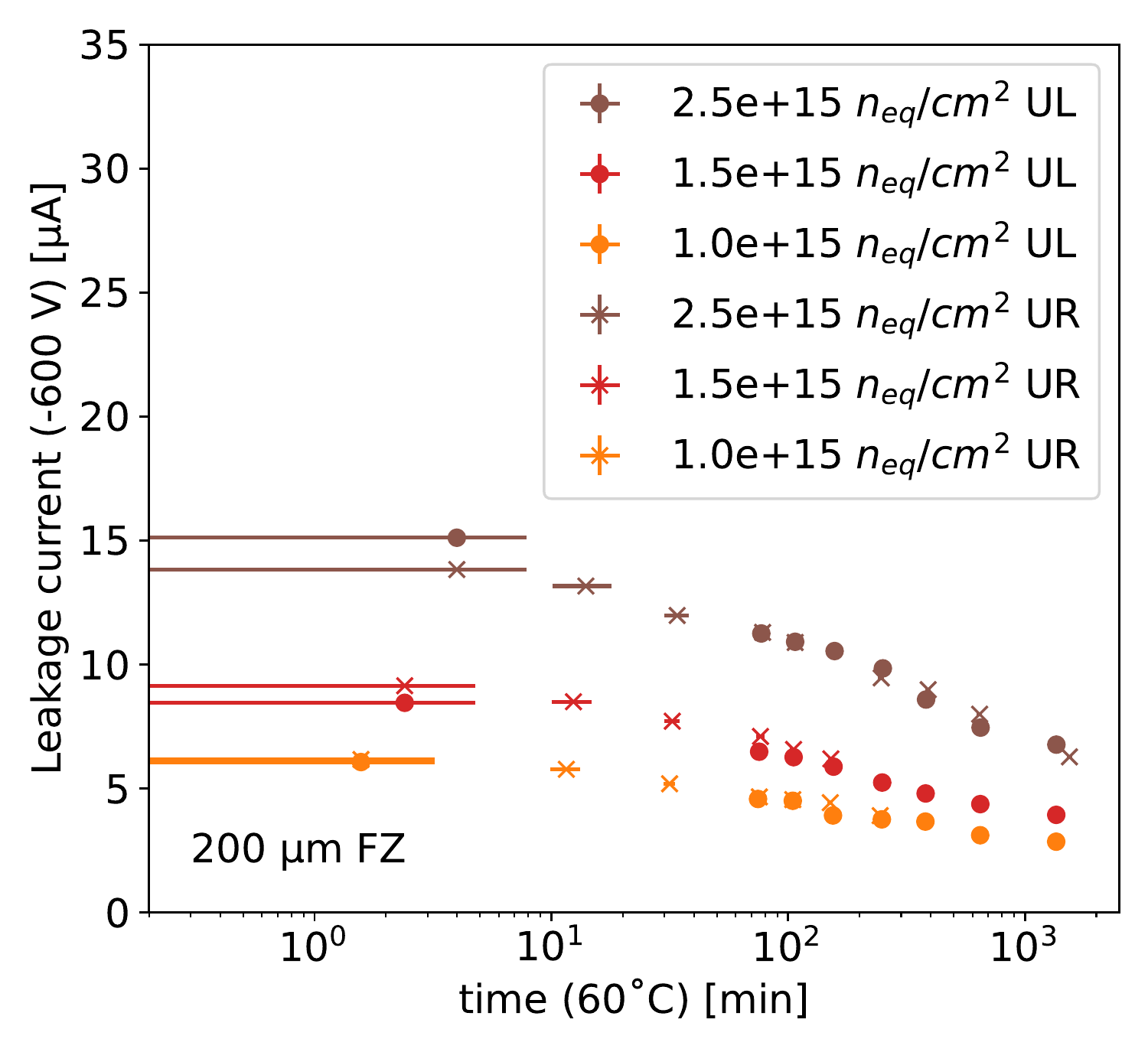}
    \includegraphics[width=0.32\textwidth]{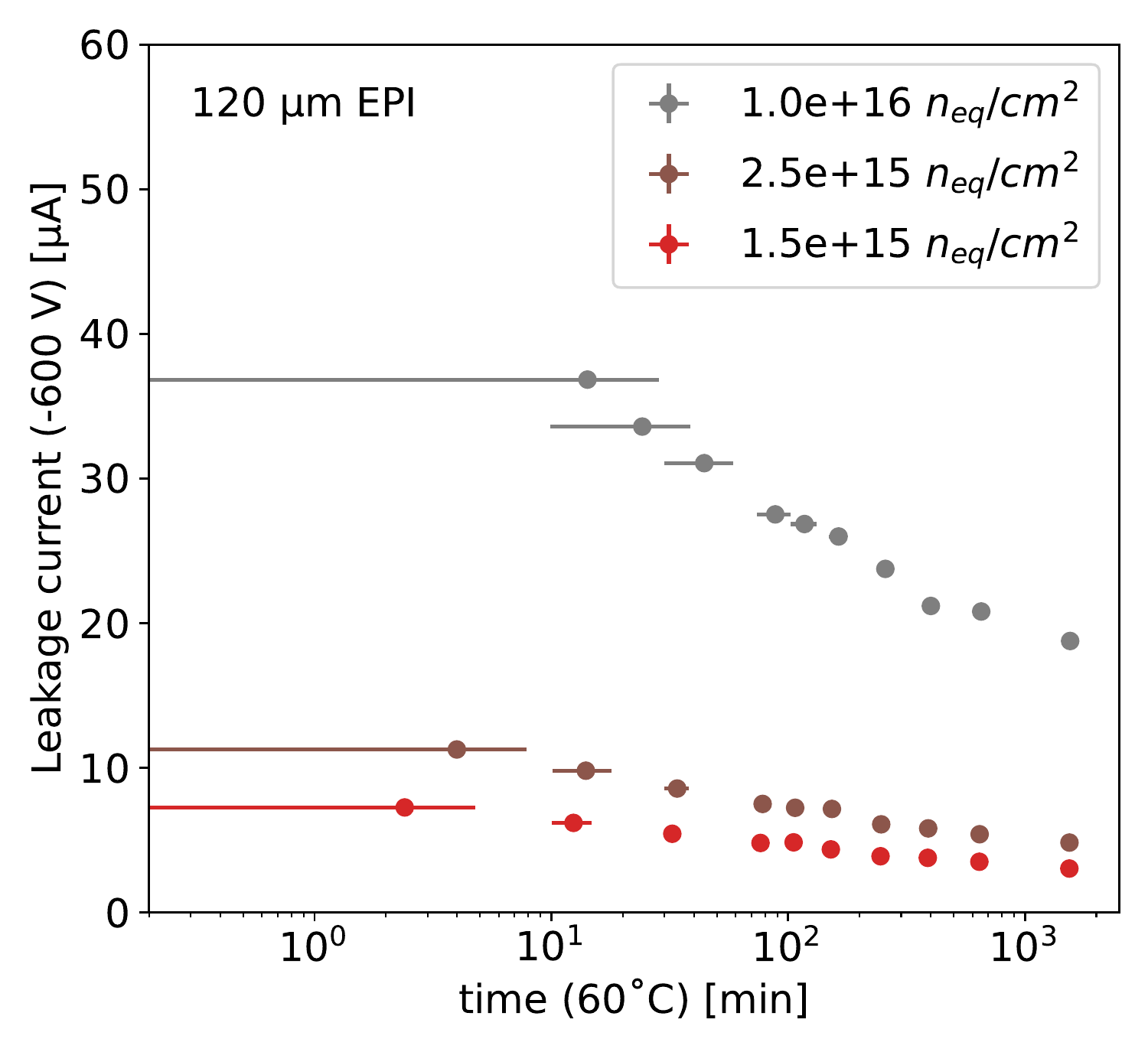}
    \includegraphics[width=0.32\textwidth]{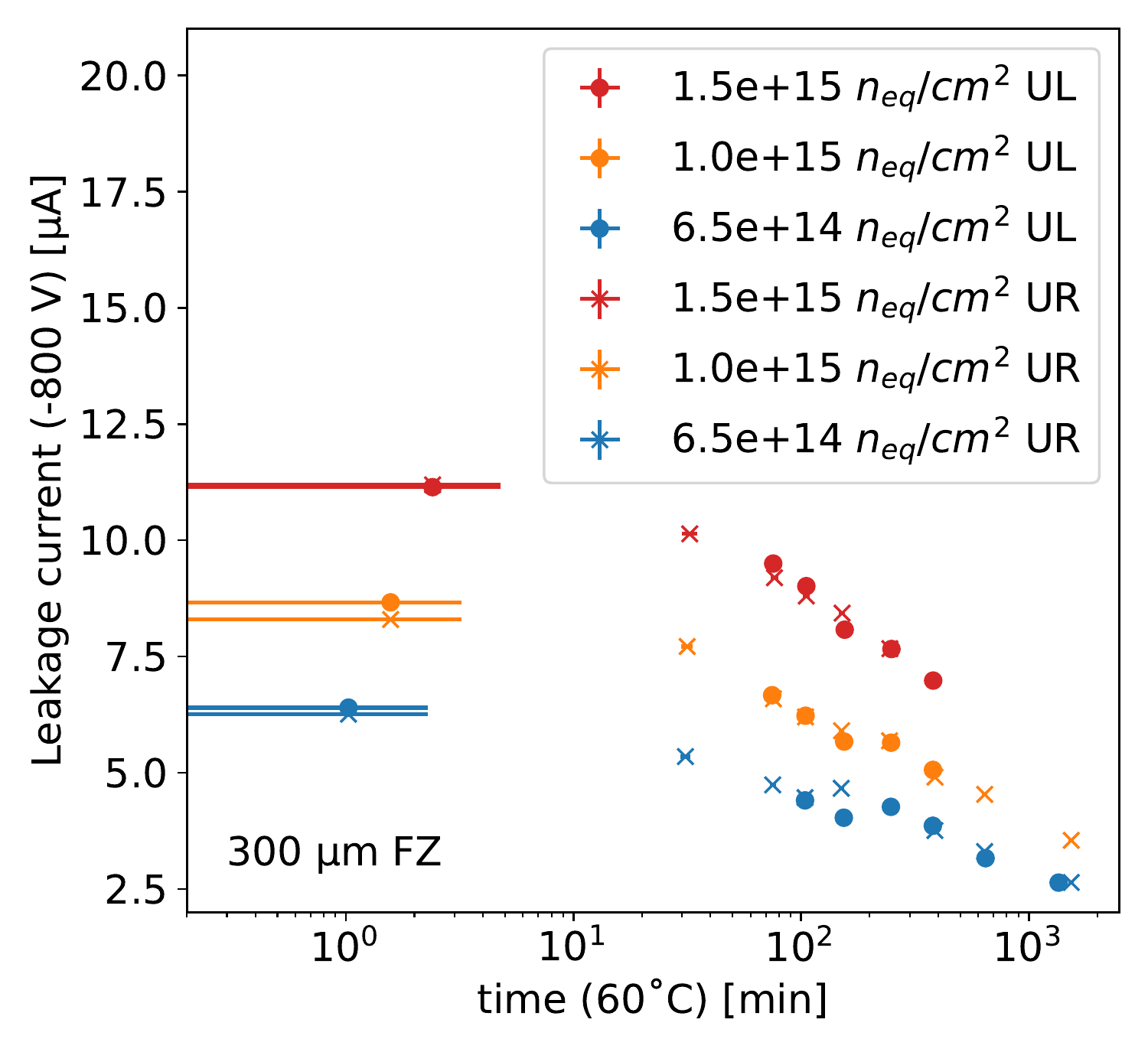}
    \includegraphics[width=0.32\textwidth]{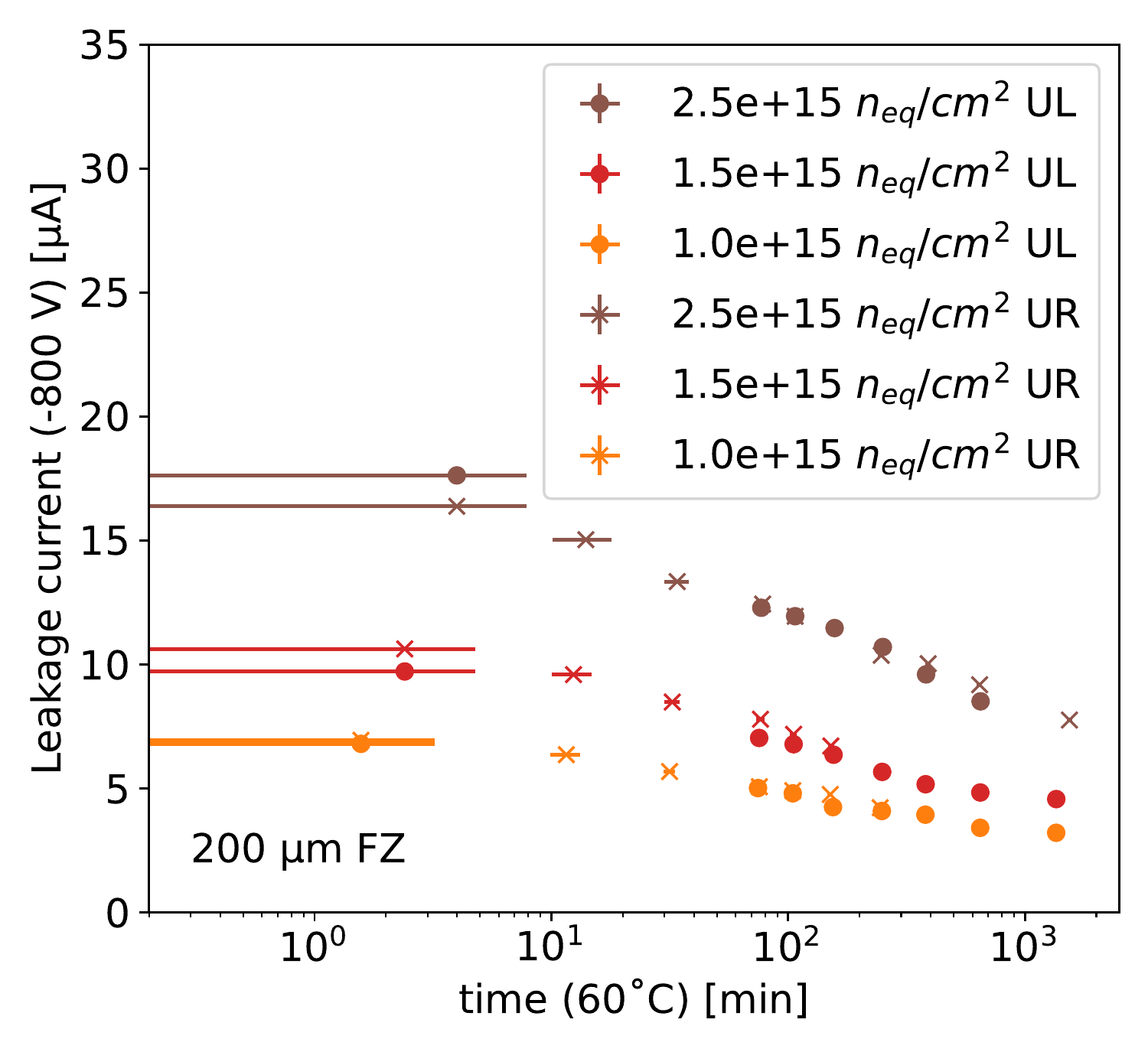}
    \includegraphics[width=0.32\textwidth]{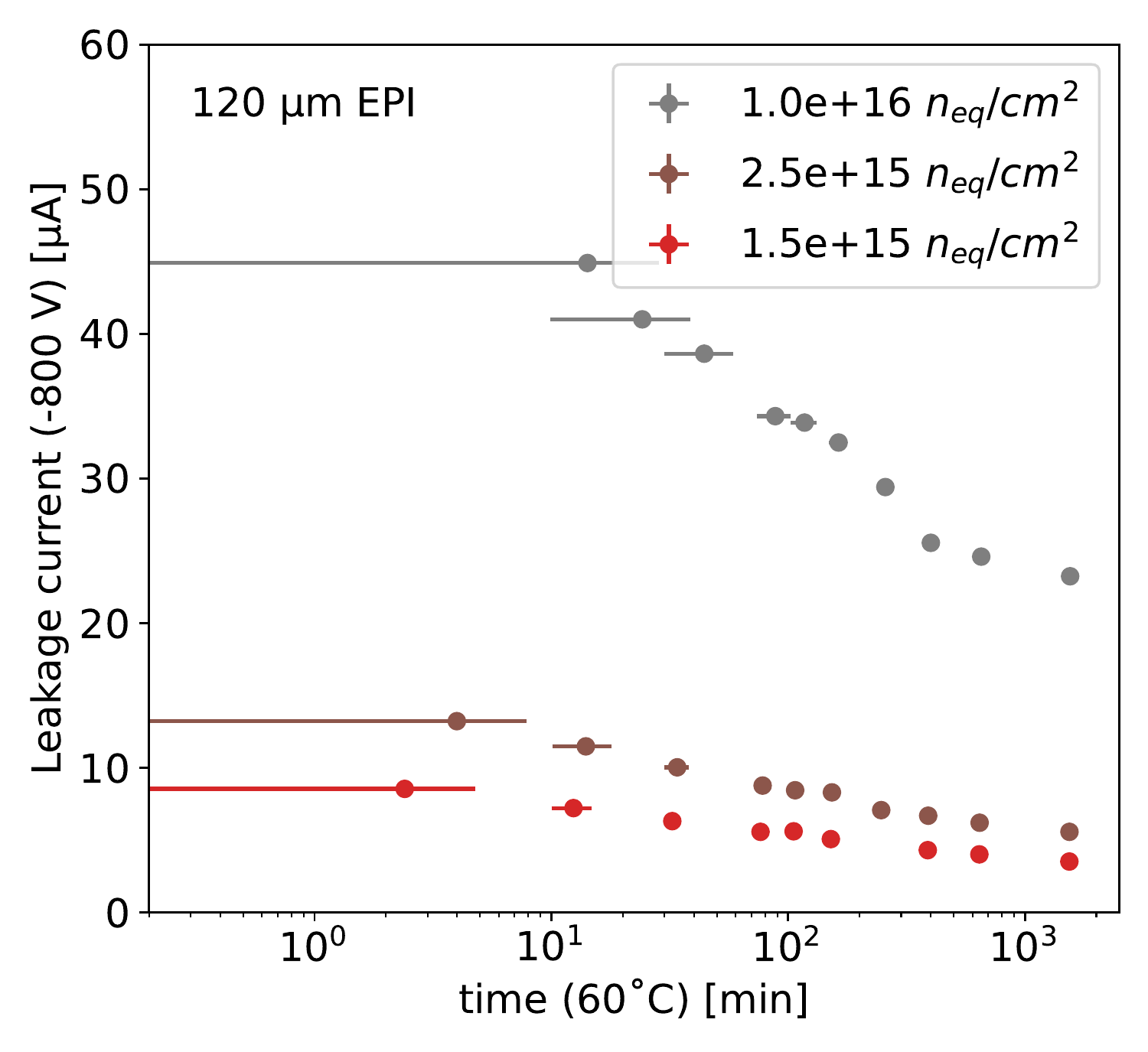}
    \includegraphics[width=0.32\textwidth]{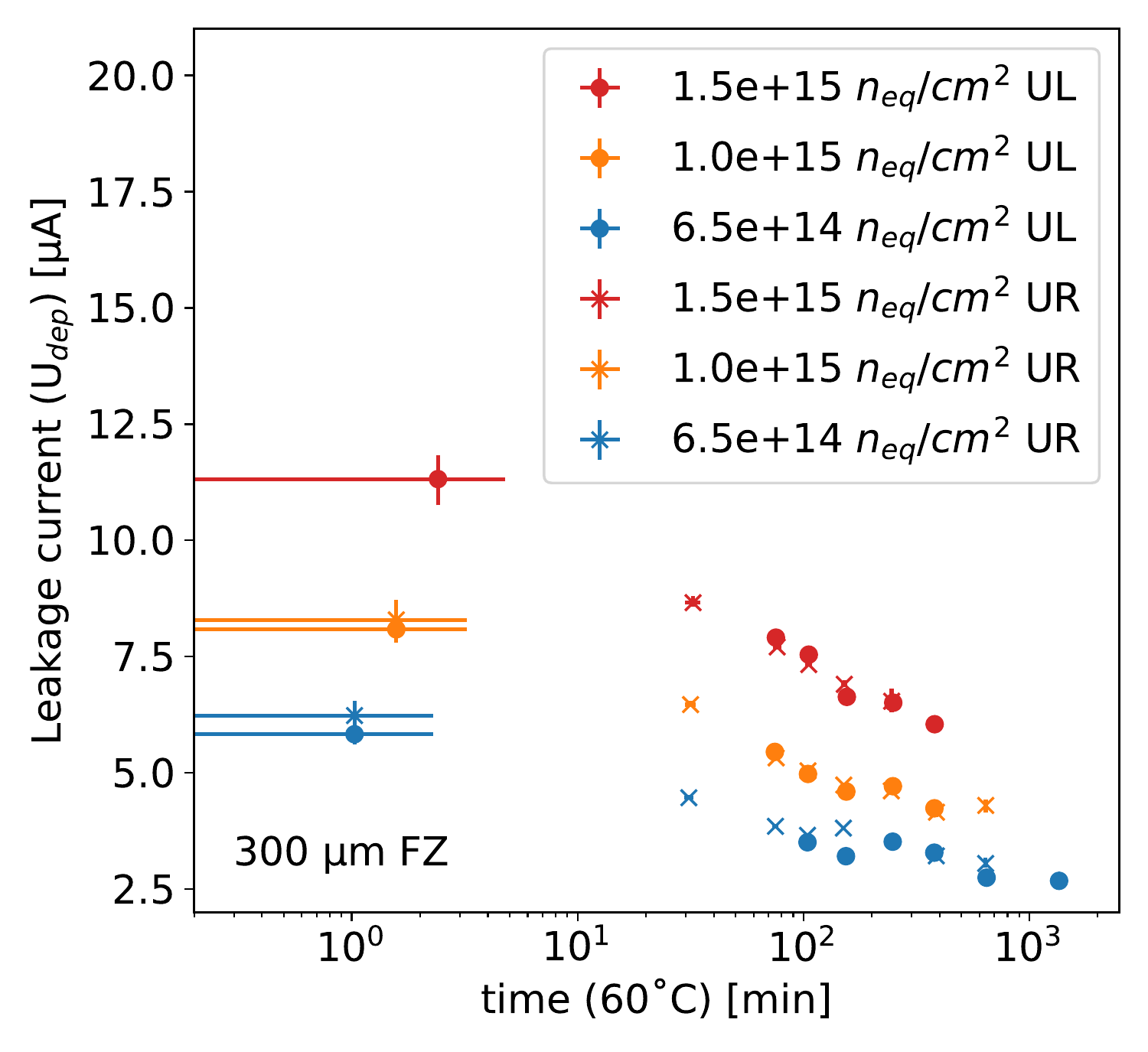}
    \includegraphics[width=0.32\textwidth]{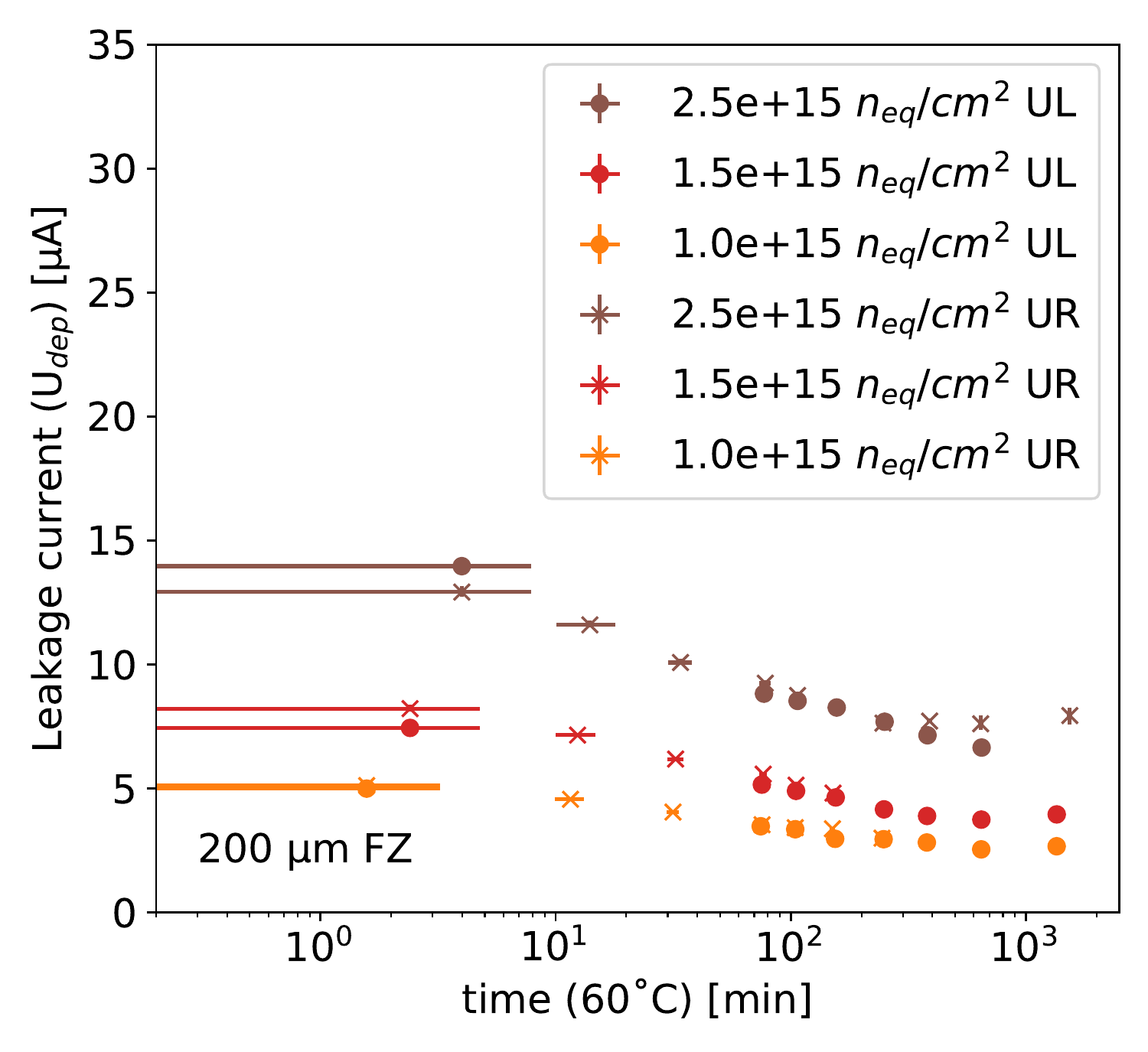}
    \includegraphics[width=0.32\textwidth]{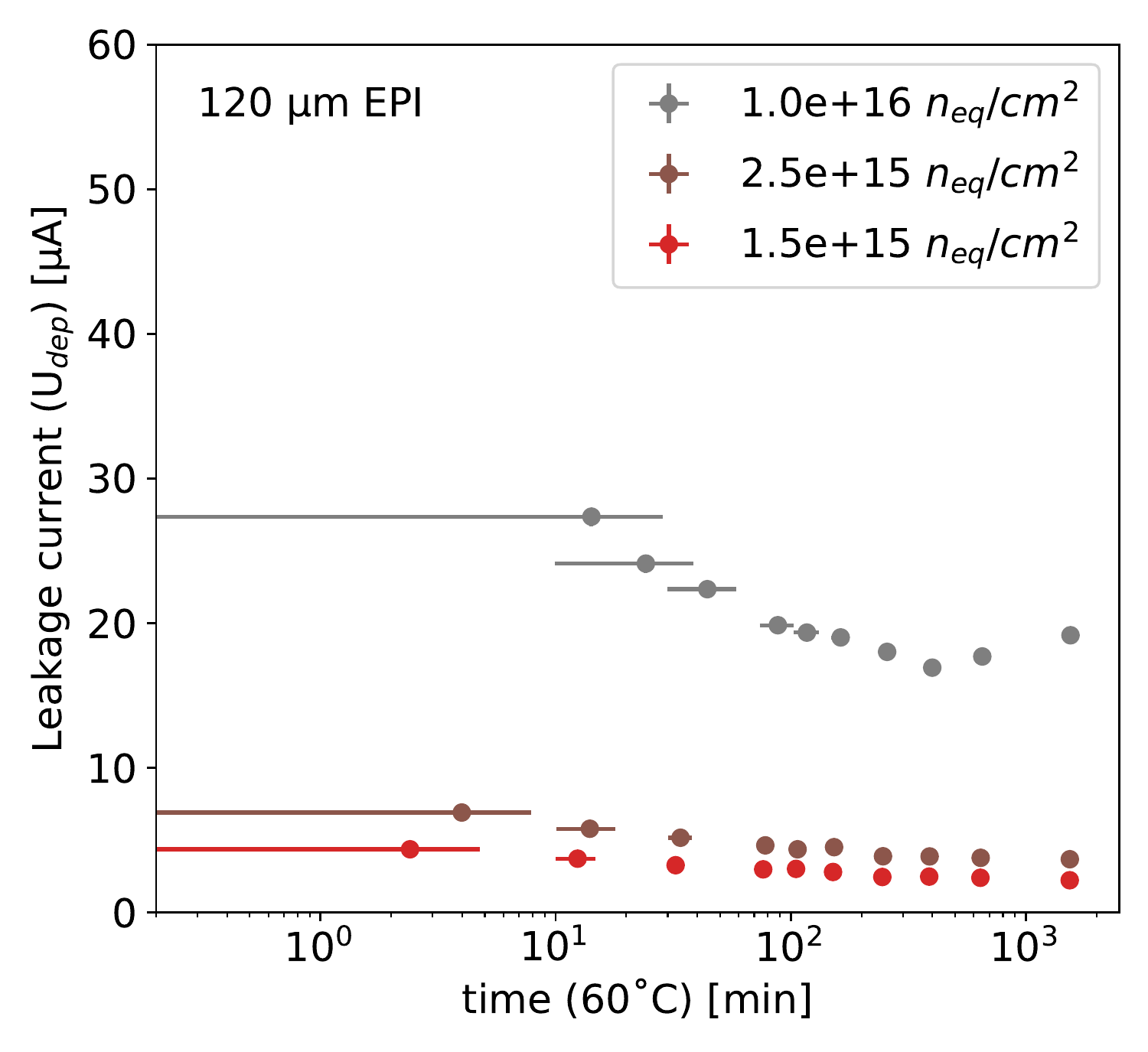}
    \caption{Leakage current for the FZ diodes with 300 (left) and 200 (middle)\micron thickness, and the EPI diodes with 120\micron thickness as a function of the equivalent annealing time at 60\degreeC, measured at -20\degreeC. The first row shows the leakage current at -600\Volt bias voltage, the second row at -800\Volt, and the third row is evaluated at depletion voltage.}
    \label{fig:leakage}
\end{figure}

The leakage current is evaluated as a function of annealing time at a fixed bias voltage of $-600\Volt$, $-800\Volt$, and at the depletion voltage. The results are shown in Figure~\ref{fig:leakage}. For all samples and fluences, the leakage current follows the expected behaviour and measurements of the UL and UR diodes are compatible for same fluences and thicknesses. The current is steadily decreasing with annealing time at a fixed bias voltage. At depletion voltage, the same behaviour can be observed until there is a slight increase at very large annealing times for high fluences, due to an increase in depletion voltage in the reverse-annealing regime shown in Figure~\ref{fig:deplvolt}. In particular at fixed bias voltage, the leakage current of the samples with 300\micron thickness shows a mild knee-like behaviour, with only small decrease in current for short annealing times. This effect disappears when evaluating the leakage current at depletion voltage, and can likely be traced back to the sensors not being fully depleted yet at either  $-600$ or $-800\Volt$.

\section{Interpretation}
\label{seq:interpretation}

\begin{figure}[htbp]
    \centering
    \includegraphics[width=0.48\textwidth]{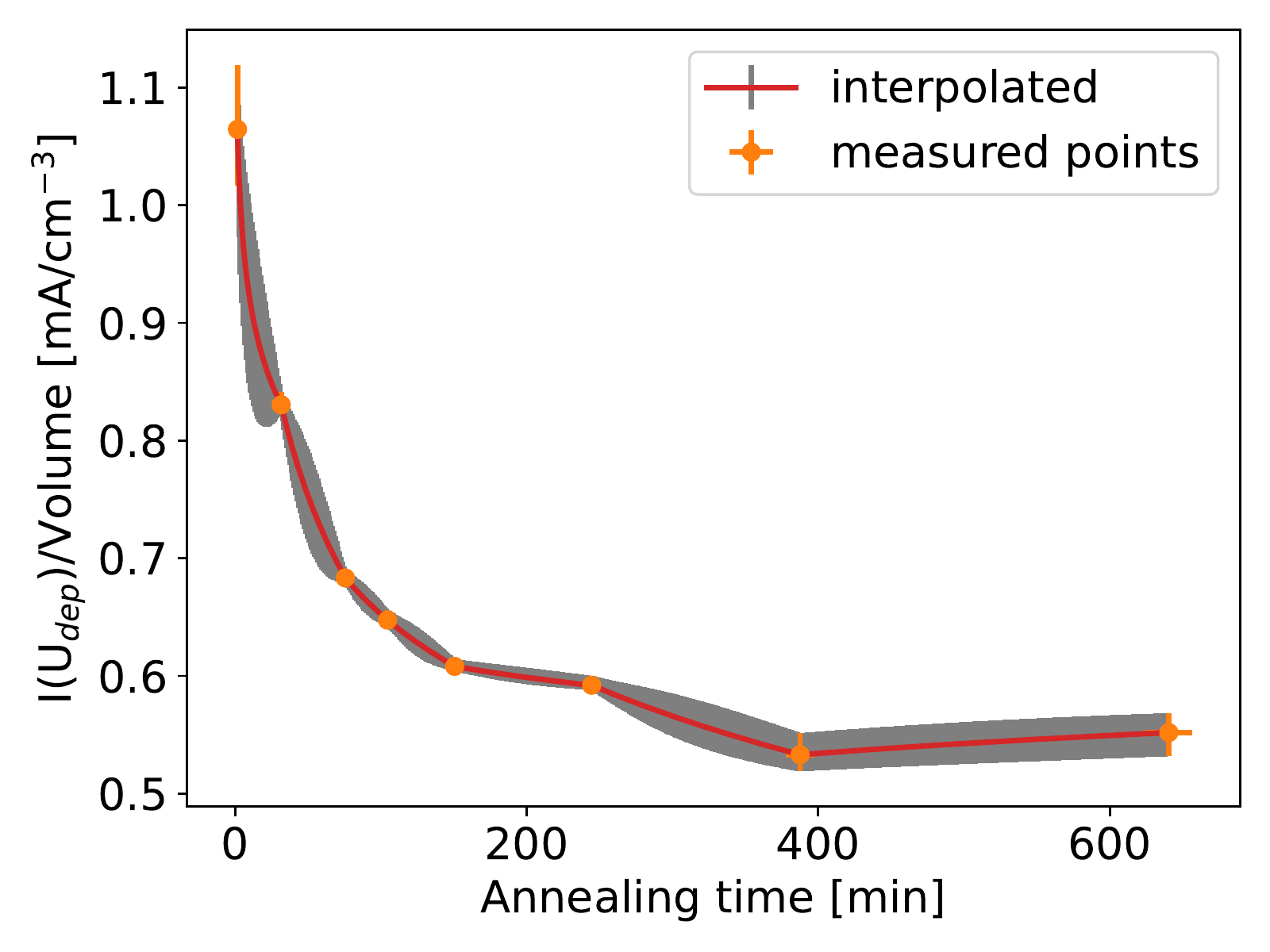}
    \caption{Interpolation of the leakage current per diode volume at depletion voltage for the FZ sample 1003 UR, with a thickness of 300\micron and a fluence of $10^{15}$\neqcm. The orange markers depict the measured points, while the gray band corresponds to the uncertainty on the interpolated curve shown in red.}
    \label{fig:interp}
\end{figure}

The acquired data is further used to investigate the annealing behaviour in terms of the current related damage rate, \Ialpha, to determine the effective doping concentration as a function of annealing time, and to parametrise the latter in terms of the Hamburg model~\cite{Moll:1999kv}. For the extraction of \Ialpha, the leakage current needs to be evaluated at different annealing times and for different fluences. For a particular annealing time, \Ialpha can be extracted by a linear fit to the leakage current at depletion voltage normalised to the diode volume, as a function of fluence. For this fit, we consider all samples for which we have measured these quantities at a certain annealing time. However, due to different time offsets, depending on the fluence, this requires an interpolation between the measured points. As we expect the leakage current to decrease linearly with the logarithm of the annealing time, we interpolate between the measured points in logarithmic space. The uncertainty assigned to the interpolation corresponds to half of the difference of the interpolated point to the nearest measured point. This contribution is added in quadrature to the uncertainty of the nearest measured point. The contribution of the two closest measured points is weighted according to the relative distance to the interpolated point in logarithmic space. An example of this interpolation and the corresponding uncertainties is shown in Figure~\ref{fig:interp}.

\begin{figure}[htbp]
    \centering
    \includegraphics[width=0.49\textwidth]{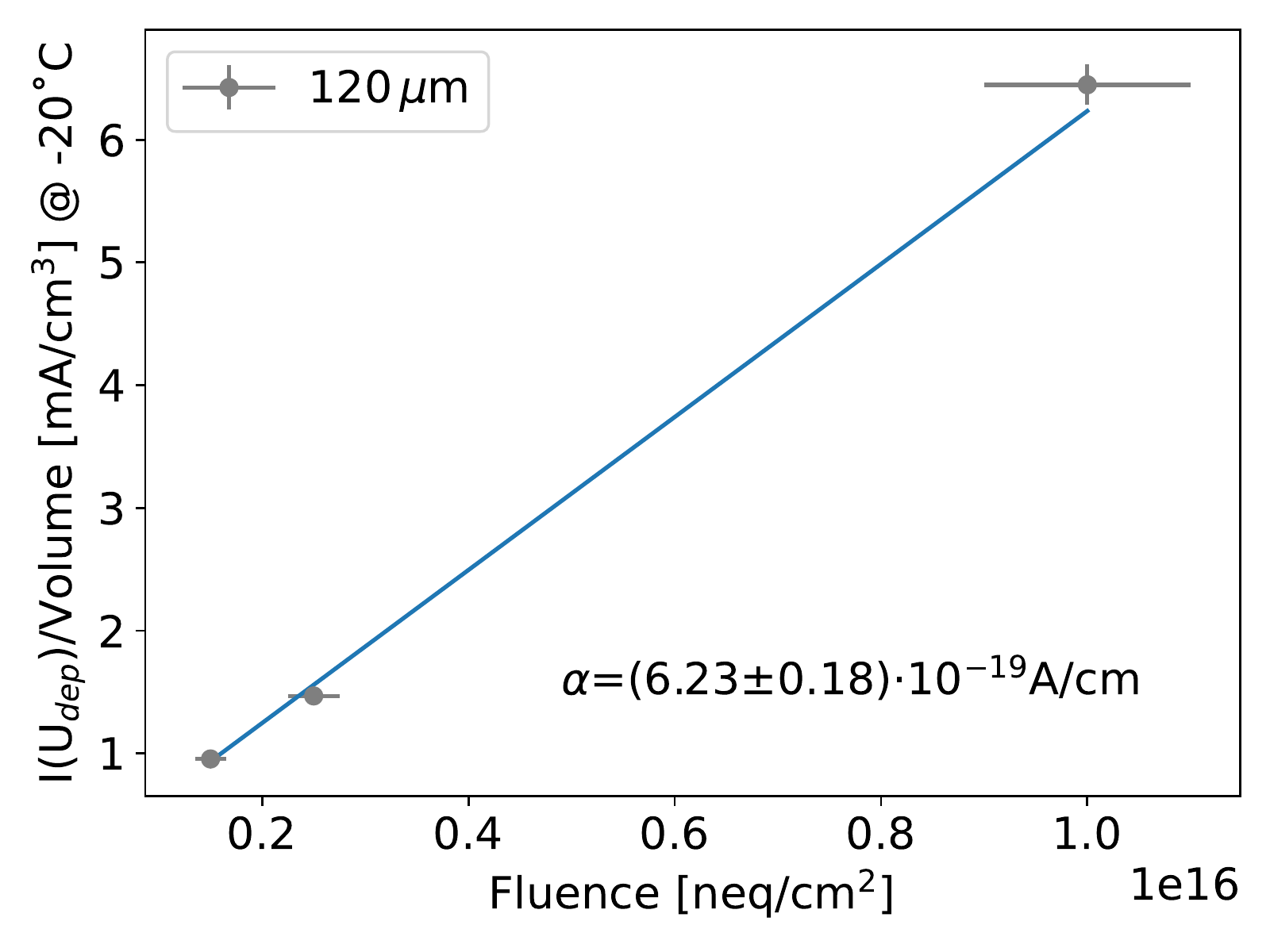}
    \includegraphics[width=0.49\textwidth]{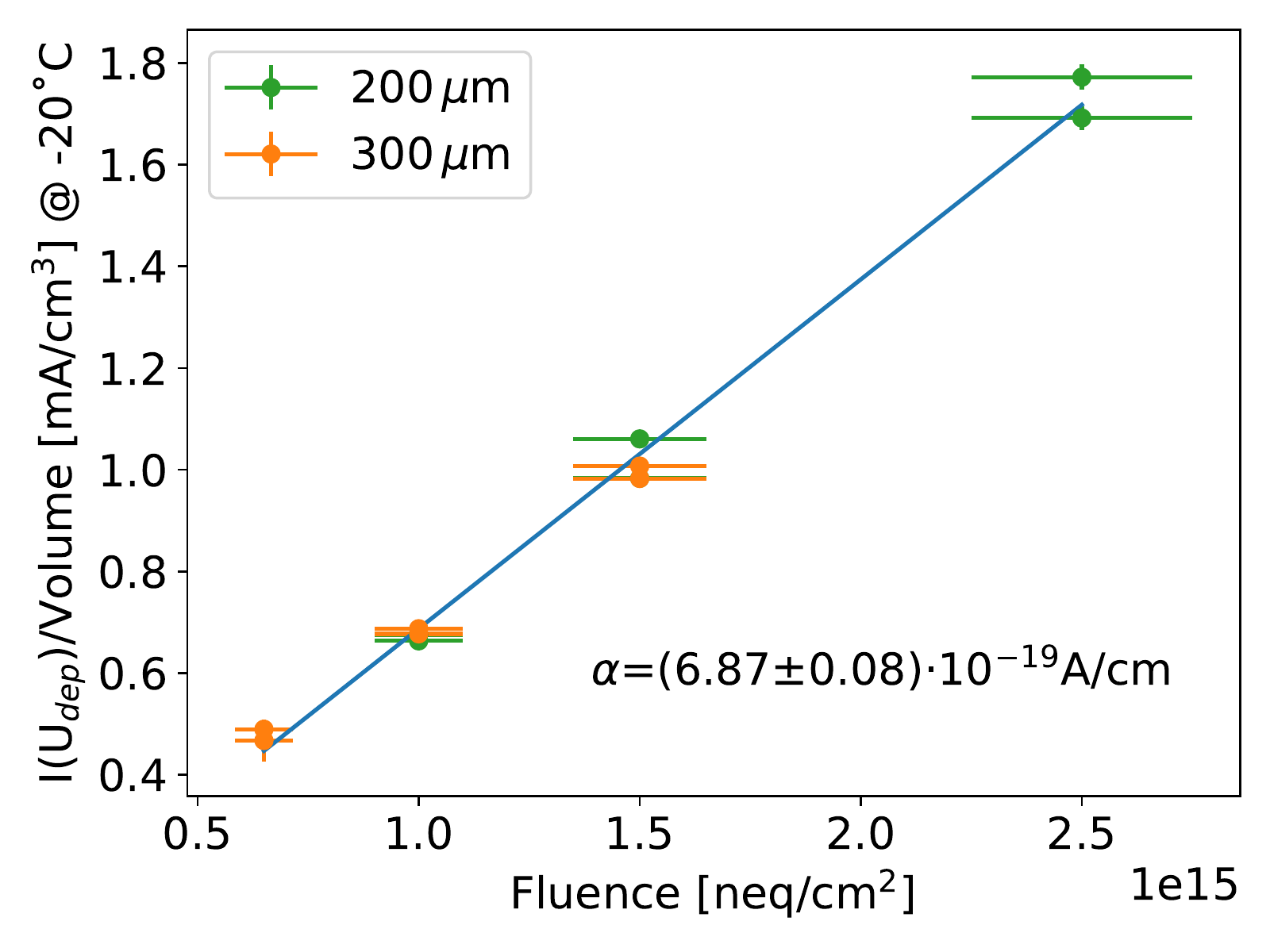}
    \caption{Leakage current at depletion voltage after 80\minutes of annealing at 60\degreeC for the EPI material (left) and the FZ material (right). The line indicates the linear fit used to extract \Ialpha, here assuming uncorrelated uncertainties in the fluences.}
    \label{fig:alpha_fit}
\end{figure}

For the extraction of \Ialpha, we select annealing times close to the measured points, and where the depletion voltage is still within the measurement range, below 1000\Volt. The linear fit is performed using an orthogonal distance regression~\cite{Boggs1989OrthogonalDR} accounting for a 10\% uncertainty on the fluence values. The exact correlation of this uncertainty between the measured points is unknown. Therefore, the results are reported assuming this uncertainty to be fully correlated and in addition treating it as fully uncorrelated.
An example of such a fit is shown for an annealing time of 80\minutes in Figure~\ref{fig:alpha_fit}.

\begin{figure}[htbp]
    \centering
    \includegraphics[width=0.58\textwidth]{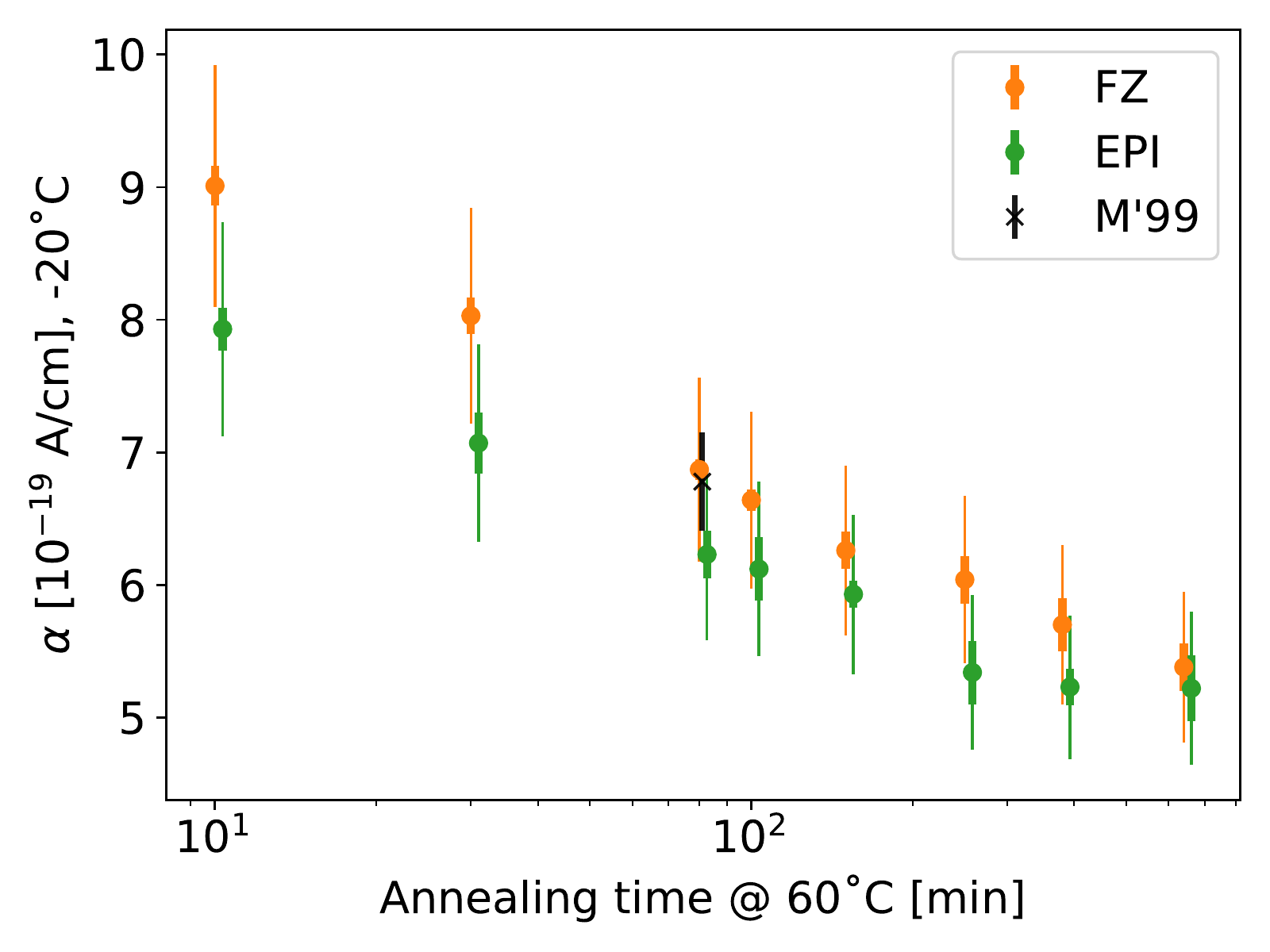}
    \caption{Current related damage rate \Ialpha as a function of annealing time. The thick error bars indicate the uncertainty on each point when assuming uncorrelated uncertainties while extracting \Ialpha, while the thin error bars represent the uncertainties when assuming the fluence uncertainty to be fully correlated across the measured points. The reference value referred to as \textit{MM'99} is taken from Ref.~\cite{Moll:1999kv} and converted to -20\degreeC.}
    \label{fig:alpha}
\end{figure}

We determine \Ialpha for the FZ and EPI samples individually. As shown in Figure~\ref{fig:alpha}, the results are mostly compatible with each other and follow the expected logarithmic falling behaviour. A slight trend towards lower values can be observed for the EPI samples, which were exposed to higher fluences. 
While we exposed the samples to a higher fluence range, the observed \Ialpha are in good agreement with the literature~\cite{Moll:1999kv}. For this comparison the latter is converted from $T_0=20\degreeC$ to $T_1=-20\degreeC$ using a conversion factor $c_I$ calculated as
\begin{equation}
    c_I = \left(\frac{T_1}{T_0}\right)^2 \exp{\frac{E_\text{ef}}{2 k_B} (1/T_0 - 1/T_1)} \text{,}
\end{equation}
with the activation energy of $E_\text{ef} = 1.21\, \text{eV}$ following the RD50 recommendations~\cite{Chilingarov:1511886}. For a conversion from 20\degreeC to -20\degreeC, this results in $c_I = 0.0169$ and a converted $\alpha = (6.8 \pm 0.4) \cdot 10^{-19}\ \mathrm{A/\cm} $ from Ref.~\cite{Moll:1999kv}.

\ 
The dependence of the depletion voltage on the annealing time can be interpreted using the Hamburg model when expressed in terms of the effective doping concentration, \Neff, which can be determined from the depletion voltage \Udep as:
\begin{equation}
    \Neff = 2 \frac{\epsilon_0 \epsilon } {q_0} \frac{\Udep}{d^2} \mathrm{,}
\end{equation}
where $\epsilon_0$ is the dielectric constant in vacuum ($8.85\cdot 10^{-14}$F/\cm), $\epsilon$ the material-dependent permittivity modifier (11.68 for silicon), $q_0$ the charge of one electron in Coulomb ($1.6\cdot 10^{-19} C$), and $d$ the diode thickness. For the non-irradiated case, the measured depletion voltages and corresponding effective doping concentrations \Neffnorad are listed in Table~\ref{tab:unirr}.

\begin{table}[h!]
    \centering
    \caption{Depletion voltage and \Neff for non-irradiated diodes.}
    \begin{tabular}{c|c|c}
    Diode thickness and material & \Udep [\Volt] & \Neffnorad [$10^{12}/\cm^{3}$] \\
        \hline
    300\micron, FZ & $263.0 \pm 1.2$ & $3.78 \pm 0.02$ \\
    200\micron, FZ & $120.0 \pm 1.5$ & $3.88 \pm 0.05$ \\
    120\micron, EPI & $41.8 \pm 0.2$ & $3.75 \pm 0.02$\\
    \end{tabular}
    \label{tab:unirr}
\end{table}

\begin{figure}[htbp]
    \centering
    \includegraphics[width=0.48\textwidth]{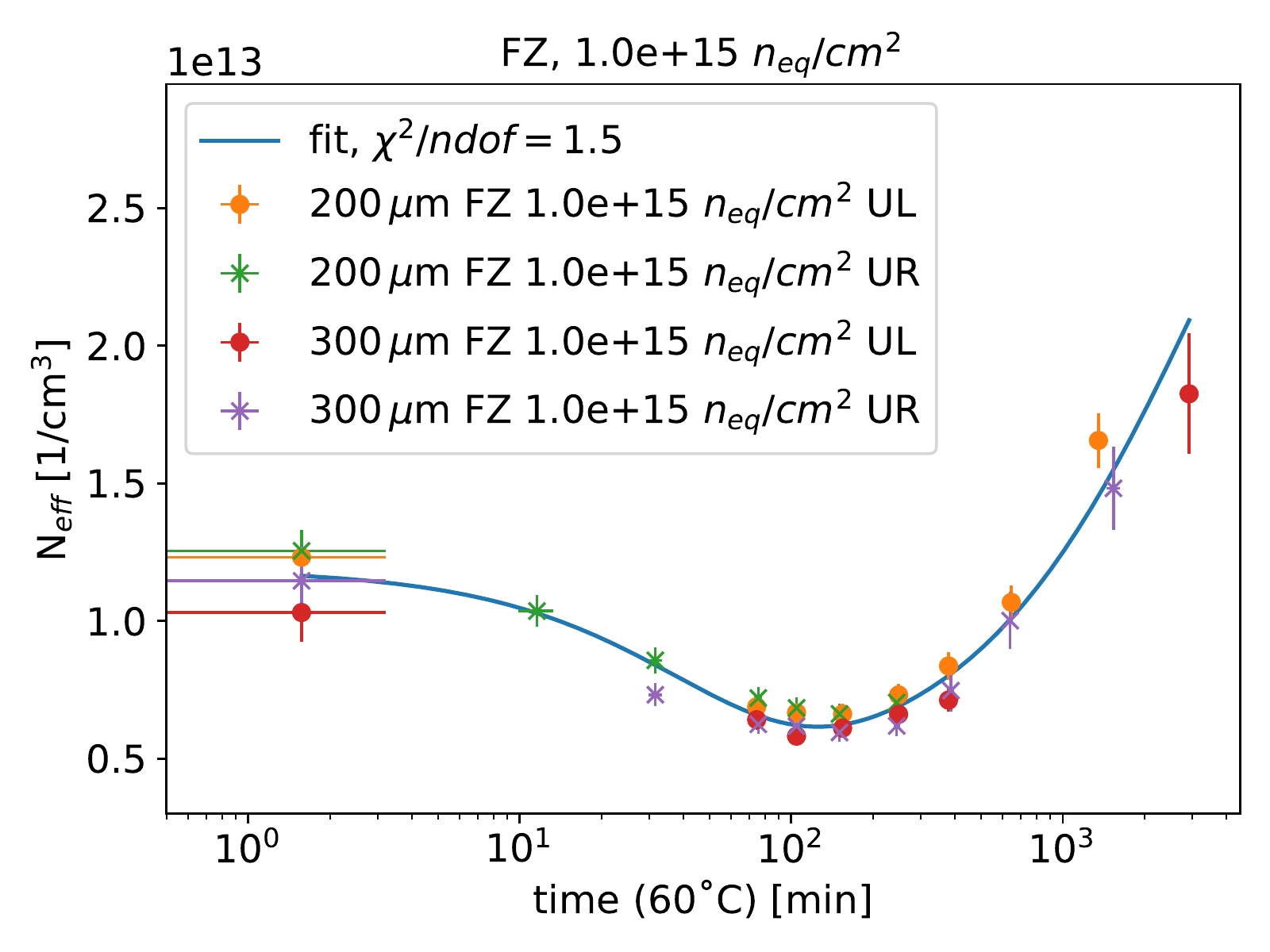}
    \includegraphics[width=0.48\textwidth]{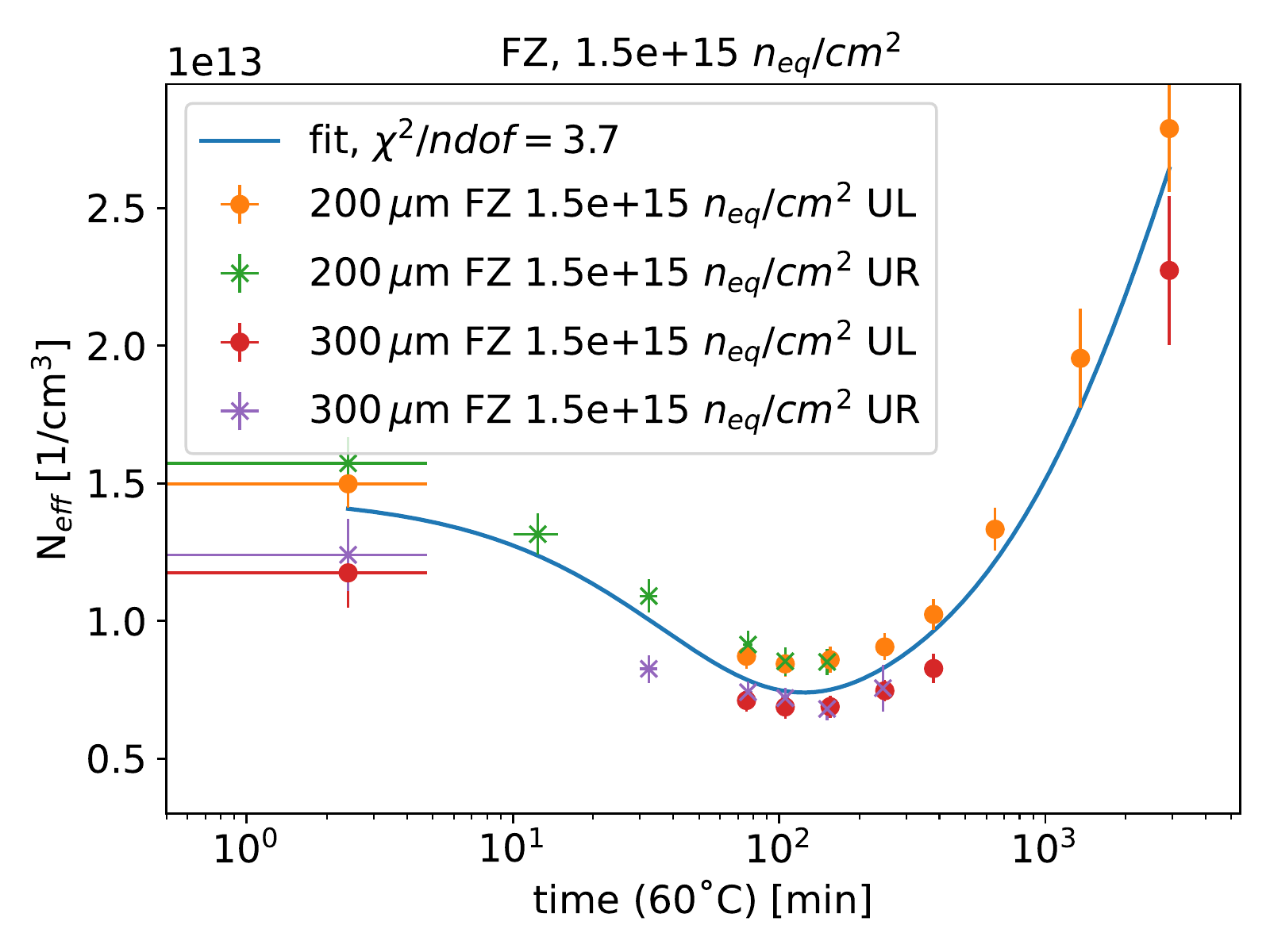}
    \includegraphics[width=0.48\textwidth]{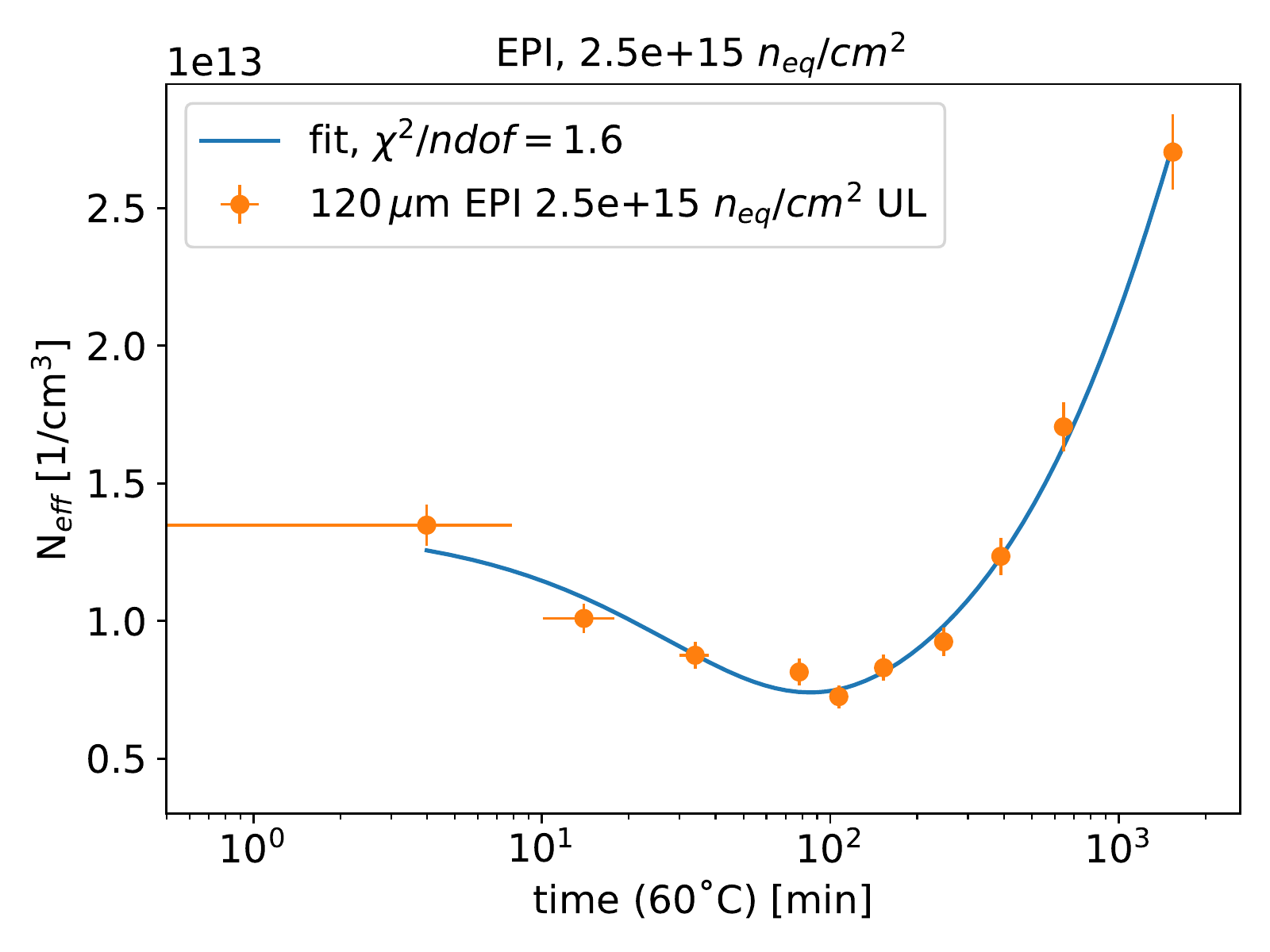}
\caption{
    Effective doping concentration as a function of annealing time fitted with the Hamburg model parameterisation for the fits with the largest $\chi^2/$ndof. 
    \label{fig:indiv_hamburg}}
\end{figure}

Three components can be identified to describe the change of \Neff due to radiation damage for equivalent annealing times of the order of minutes or longer at 60\degreeC~\cite{Lindstrom:1999mw,Moll:1999kv}: 

\begin{eqnarray}
N_A & = &  g_a \Phi \exp(-t/\tau_a) \\
N_C & = & N_{C0}(1 - \exp(-c\Phi)) + g_c\Phi \\
N_Y & = & g_y \Phi \left(1- \frac{1}{1+t/\tau_Y}\right)
\end{eqnarray}

where $\Phi$ is the fluence, and $t$ is the annealing time. 
The term $N_A$ describes the beneficial annealing effect, reducing the depletion voltage with time, starting from an effective doping density that is proportional to the fluence by $g_a$ and falling with a time constant $\tau_a$. The term $N_C$ represents a material dependent stable damage (initial donor or acceptor removal), which can be parametrised by two fluence dependent terms, using the constants $N_{C0}$ , $c$ and $g_c$. 
The term $N_Y$ models the reverse annealing described by an upper limit $g_y$ and a time constant $\tau_Y$. 
Given the high fluences and the high resistivity material, we approximate $N_C$ here as~\cite{Moll:2640820}:
\begin{equation}
N_C = g_c\Phi
\end{equation}
Then, \Neff after radiation and during annealing can be expressed as a function of fluence and time as:
\begin{equation}
    \Neff(\Phi,t) = N_A(\Phi,t) + N_C(\Phi) + N_Y(\Phi,t) + \Neffnorad
\end{equation}
For this study, we consider $g_a$, $\tau_a$, $g_c$, $g_y$, and $\tau_Y$ free parameters.
When extracting them, the EPI and FZ material should be considered separately, since they might show different behaviour. As it has been shown that for certain materials, a linear dependence on the fluence can be violated~\cite{Lindstrom:2002gb}, we consider samples of different fluences separately, but fit samples of same fluence and material simultaneously. The uncertainties on the depletion voltage are considered uncorrelated between the points during this fit. The uncertainty on the annealing offset is considered correlated across all points from the same sensor, and becomes non-negligible in particular for the EPI samples. Here, the uncertainty is assigned by repeating the fit for all annealing offset variations. We do not restrict the parameters to be positive. 
Overall, the fits with the Hamburg model provide a good description for each fluence individually. The minimum $\chi^2$ divided by the degrees of freedom is generally well below one, with the exception of three fits shown in Figure~\ref{fig:indiv_hamburg}. The larger $\chi^2$ for the FZ material is mostly caused by an offset between different measured samples. This offset can, at least in parts, be explained by the uncertainty in the initial fluence for each sample of up to 10\%, and possibly small deviations of the effective thickness with respect to the nominal thickness. While the observed offsets introduce a larger uncertainty in the constant damage factor $g_c$, they do not have a significant effect on the time constants or the constants $g_a$ and $g_y$. The larger $\chi^2$ for the EPI sample shown in Figure~\ref{fig:indiv_hamburg} seems to be a fluctuation rather than a systematic effect.

The total uncertainty on the annealing time at minimum depletion voltage, $t(U_\text{dep,min}$), is determined using the full fit correlation matrix together with the impacts of the individual fit parameters on this quantity, since the degree of correlation between the parameters is non-negligible. For all extracted parameters, the uncertainties from the annealing offset and the parameter uncertainties are added in quadrature.

\begin{figure}[htbp]
    \centering
    \includegraphics[width=0.48\textwidth]{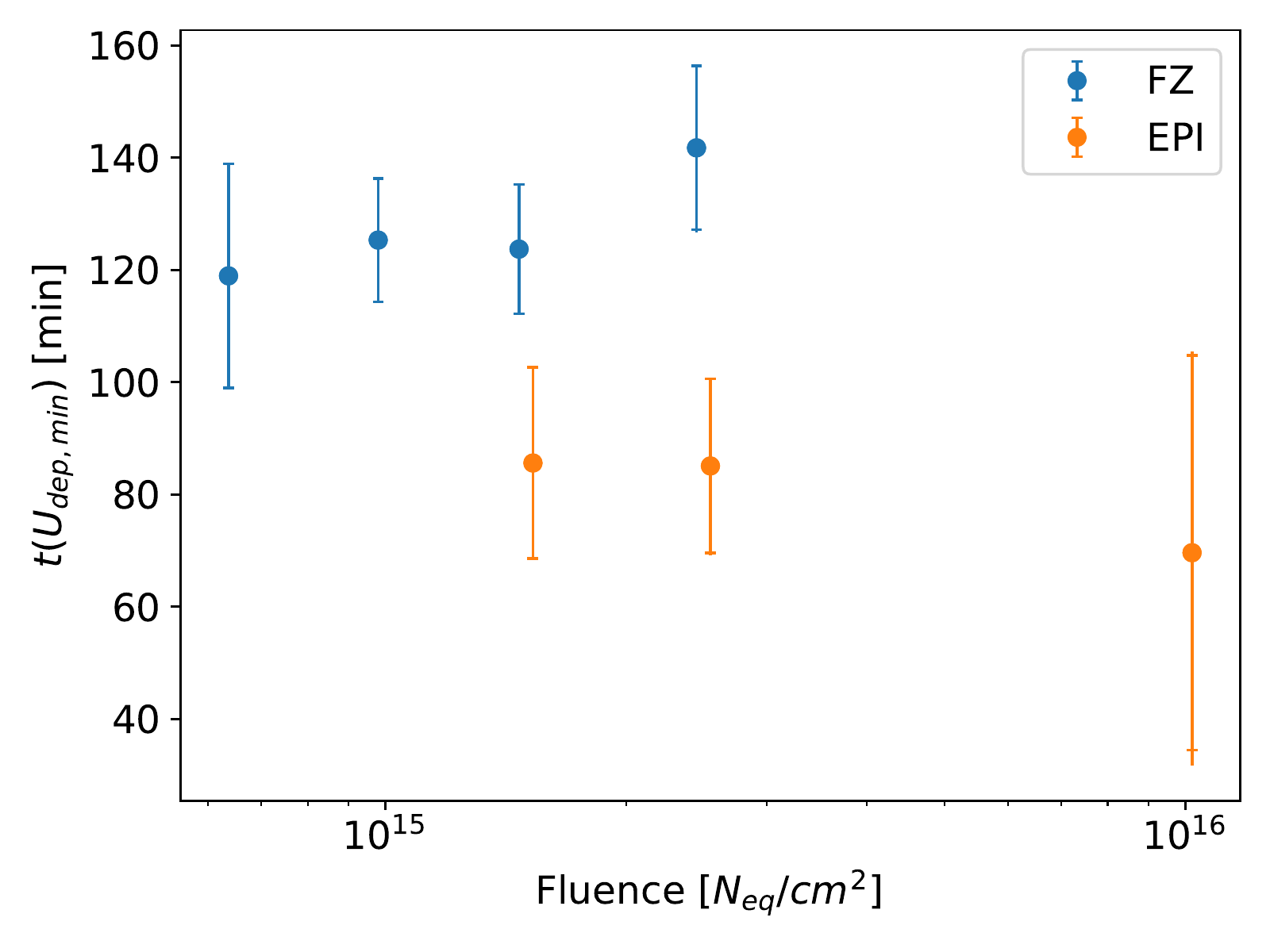}
    \includegraphics[width=0.48\textwidth]{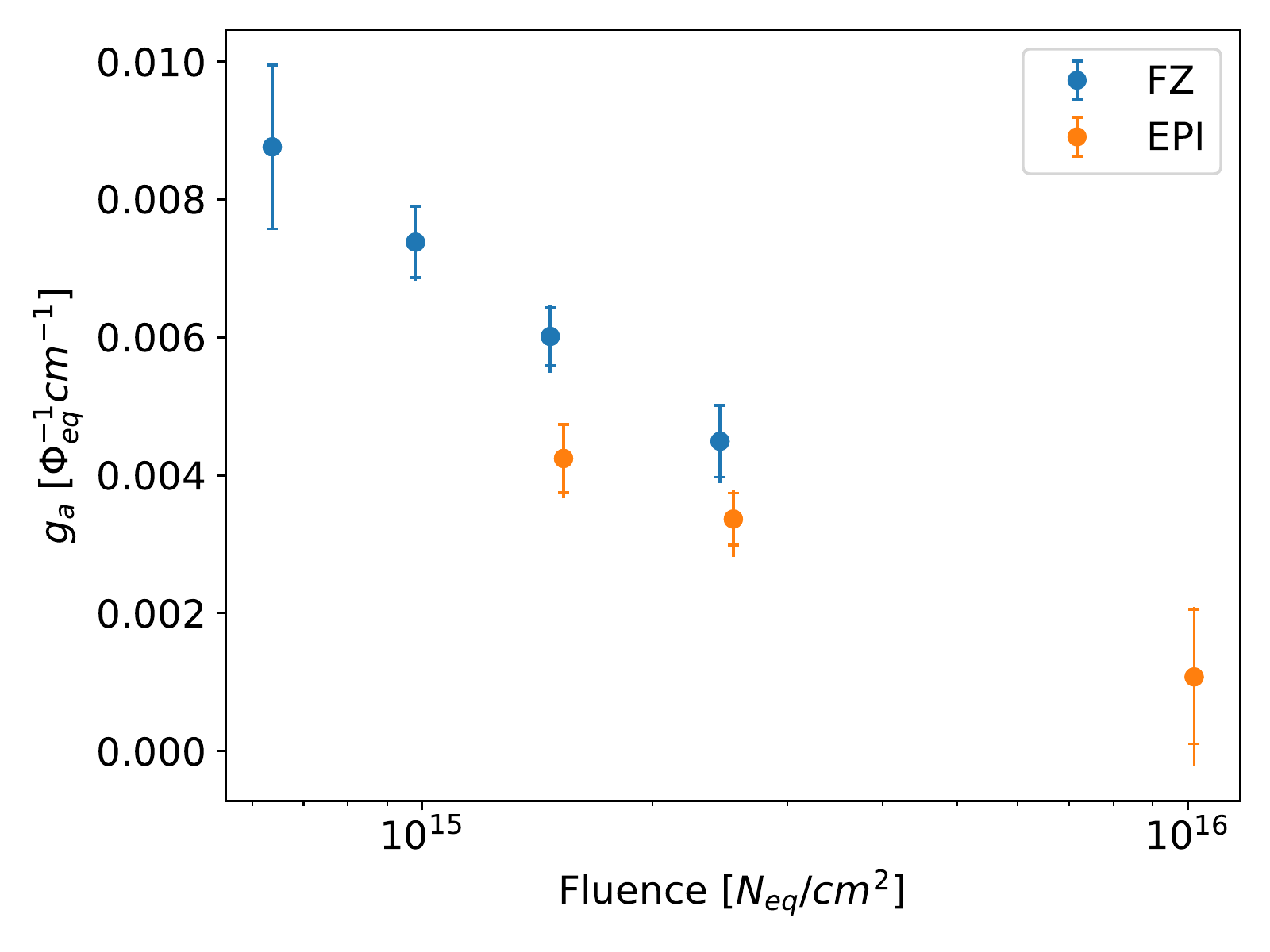}
    \includegraphics[width=0.48\textwidth]{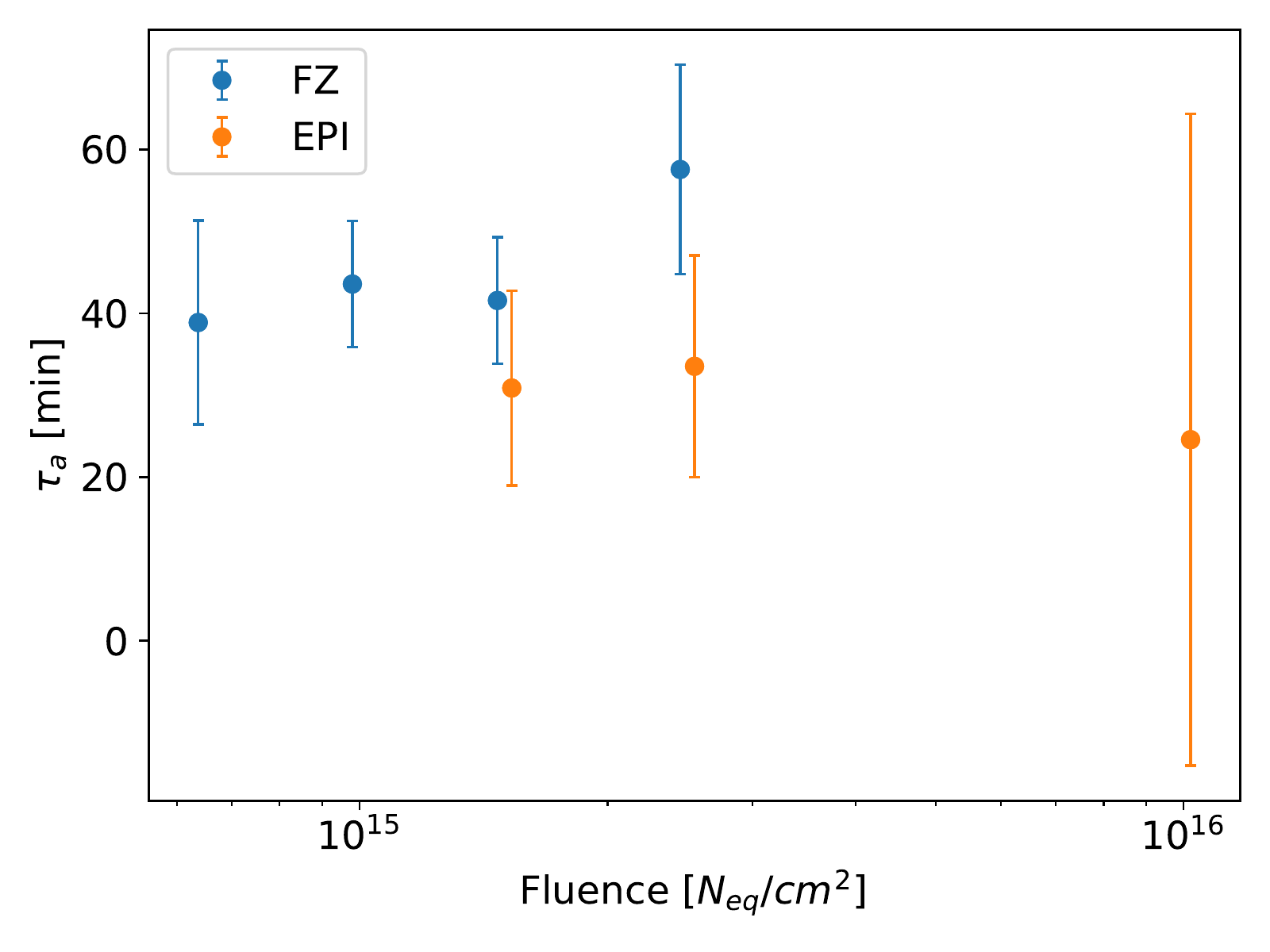}
    \includegraphics[width=0.48\textwidth]{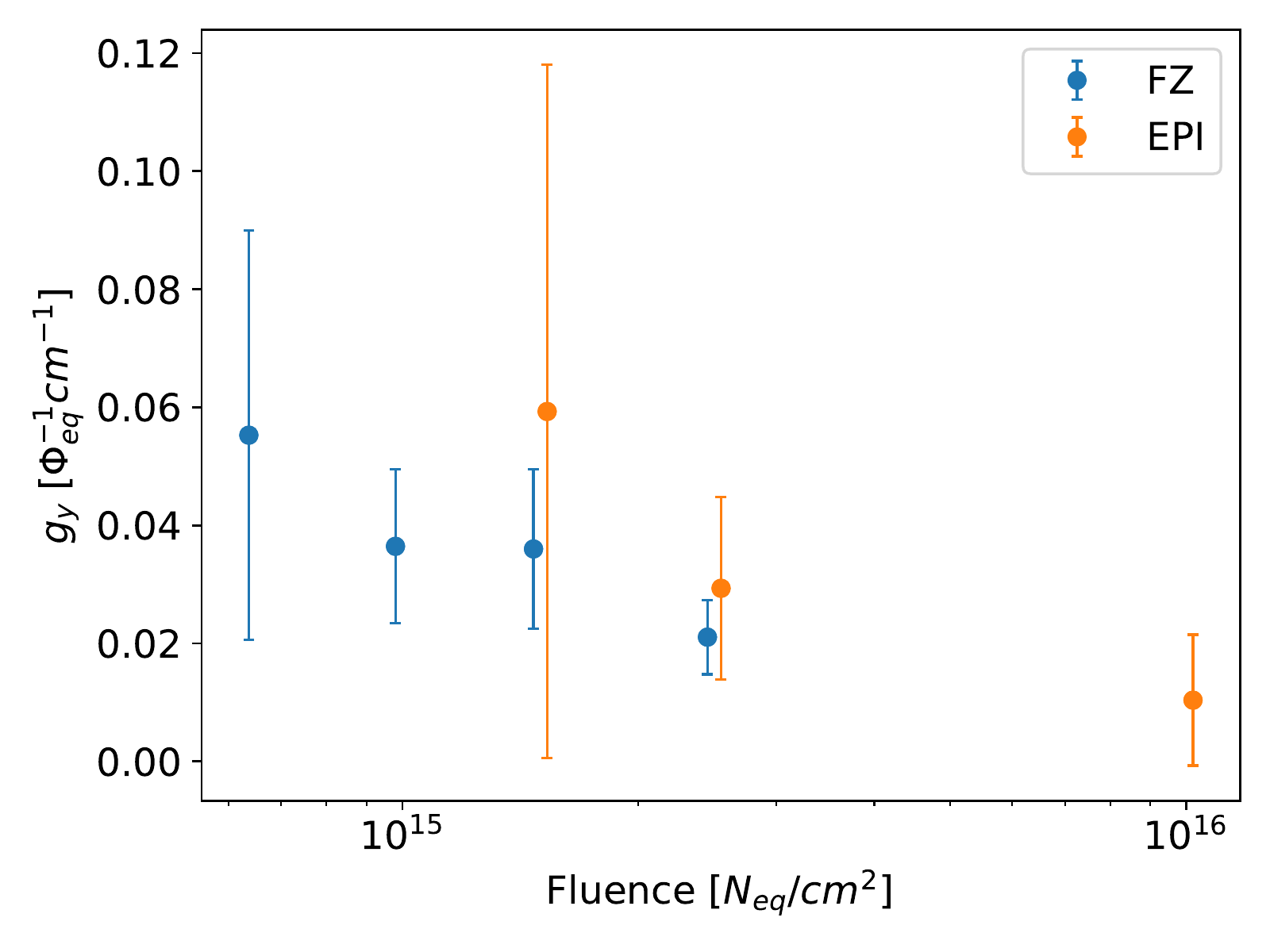}
    \includegraphics[width=0.48\textwidth]{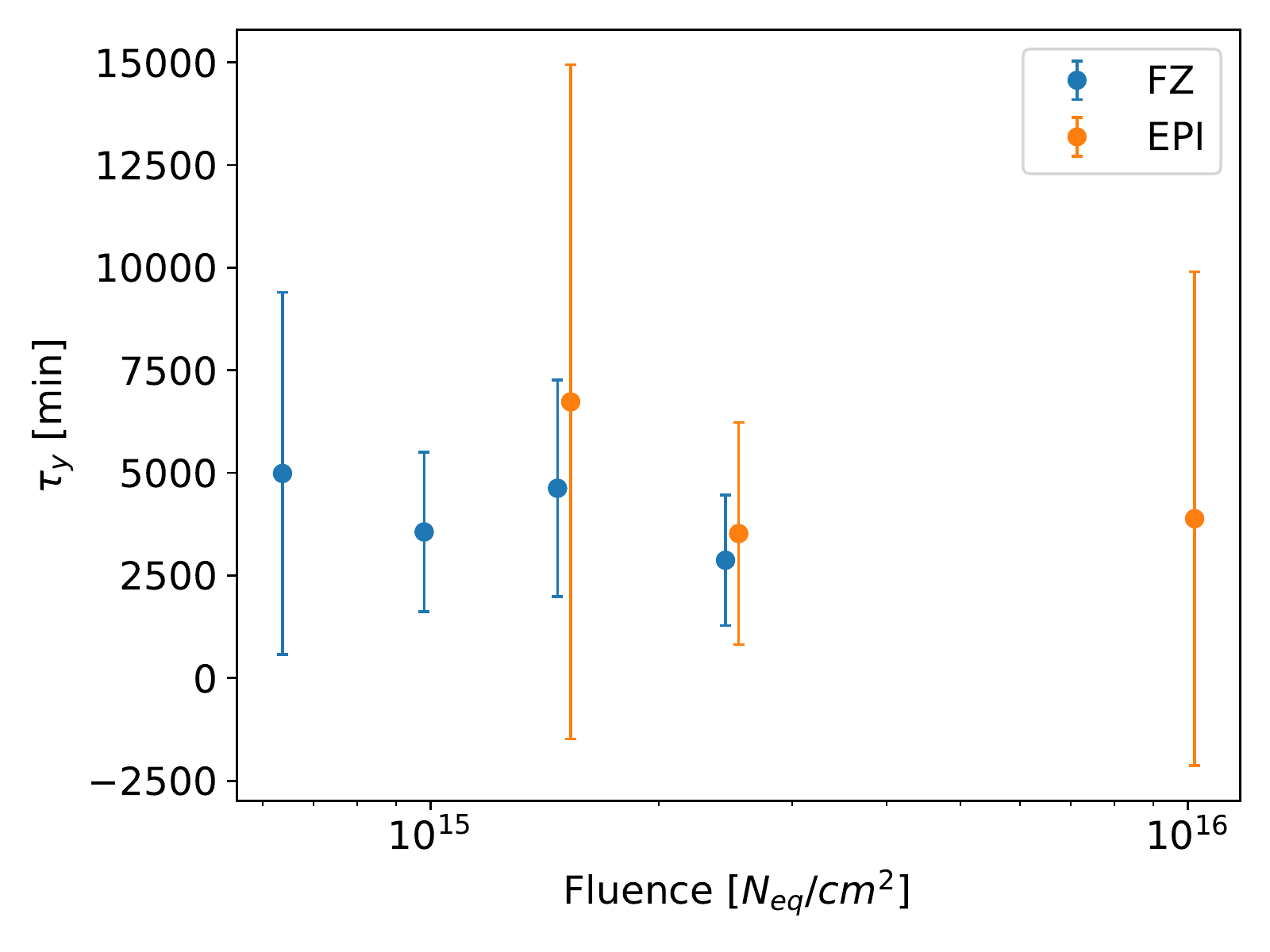}
    \includegraphics[width=0.48\textwidth]{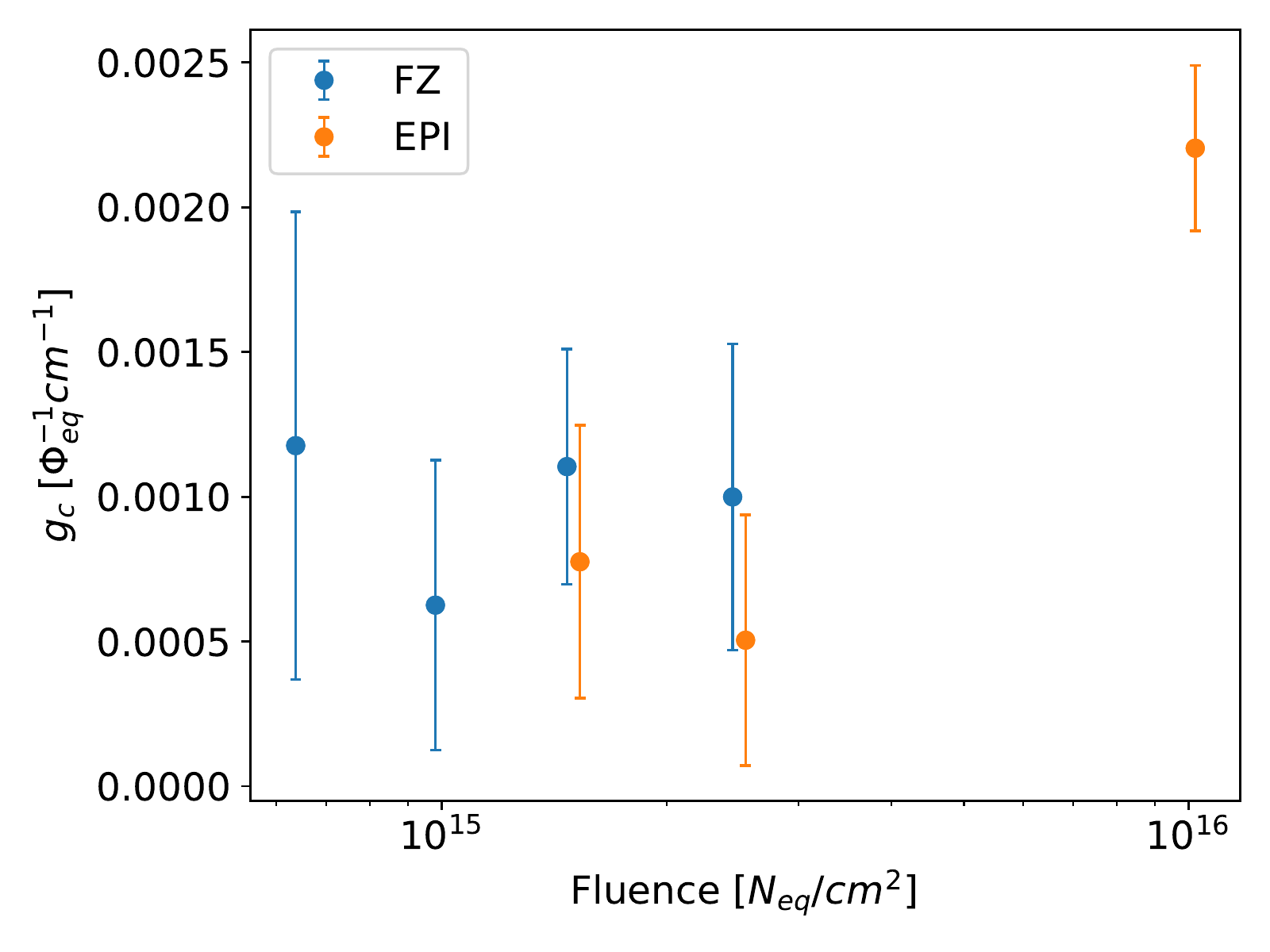}
    \caption{Fitted parameters of the Hamburg model. The inner error bars represent the fit uncertainty. The outer error bars include the uncertainty from a variation of the annealing offsets during the irradiation. These are only visible in the first row. Small offsets on the x-axes have been added for better readability, but do not indicate any differences in fluence.}
    \label{fig:hamburg}
\end{figure}

The resulting extracted parameters are shown in Figure~\ref{fig:hamburg}. 
The annealing time at which the minimum depletion voltage is reached seems consistently higher for the FZ material with 130\minutes compared to the EPI material, where values of about 80\minute seem to be preferred. A similar value for $t(U_\text{dep,min})$ was also extracted for the p-type FZ material for the ATLAS tracker upgrade~\cite{LDiehl_ptype}. Furthermore, the FZ material here shows a slight increase of $t(U_\text{dep,min})$ with the fluence.  In particular at high fluences, the uncertainty on the annealing offset starts to contribute to the total uncertainty. 
The parameter $g_a$ exhibits a consistent decrease with the fluence, common for FZ and EPI material. Also here, the annealing offset uncertainty starts to contribute at high fluences. While this increases the uncertainty, the decrease in value does not match the expectation from studies performed at lower fluences. However, the absolute value at the lowest fluence point of this study is consistent with high fluence points from Refs.~\cite{Lindstrom:1999mw,Moll:1999kv}. 
Also the parameter $g_y$ seems to show a slight decrease with fluence, even though it is much less pronounced. With the exception of the highest fluence, $g_c$ seems consistent for different materials and fluences, and the constant damage seems to scale 4--10 times less with the fluence than observed for n-type silicon~\cite{Lindstrom:1999mw,Moll:1999kv}.
The time constants $\tau_a$ and $\tau_Y$, describing the beneficial or reverse annealing, respectively, are consistent within uncertainties for different materials and fluences. 
The data indicates a decrease of the damage constants $g_a$ and $g_y$ with the fluence in contrast to a constant behaviour observed in previous measurements at lower fluences. 
 It has to be noted that the maximum annealing time studied here and limitations in the maximum bias voltage prevent  us from reaching and studying the reverse-annealing plateau. This introduces large uncertainties in and large correlations between the reverse-annealing parameters in particular, but also affects the other parameters.

The extraction of the depletion voltage from CV data is known to have a strong frequency dependence for highly irradiated sensors~\cite{FdepCampbell}. Here, the depletion voltage decreases linearly with the logarithm of the frequency, such that higher depletion voltages can be expected if the measurements were to be repeated at lower measurement frequencies. However, the choice of the frequency is to some extent arbitrary. To introduce the fluence dependence observed in particular with respect to $g_a$, the frequency dependence of the capacitance measurement would need to be a function of fluence and possibly annealing time.



\begin{figure}[htbp]
    \centering
    \includegraphics[width=0.7\textwidth]{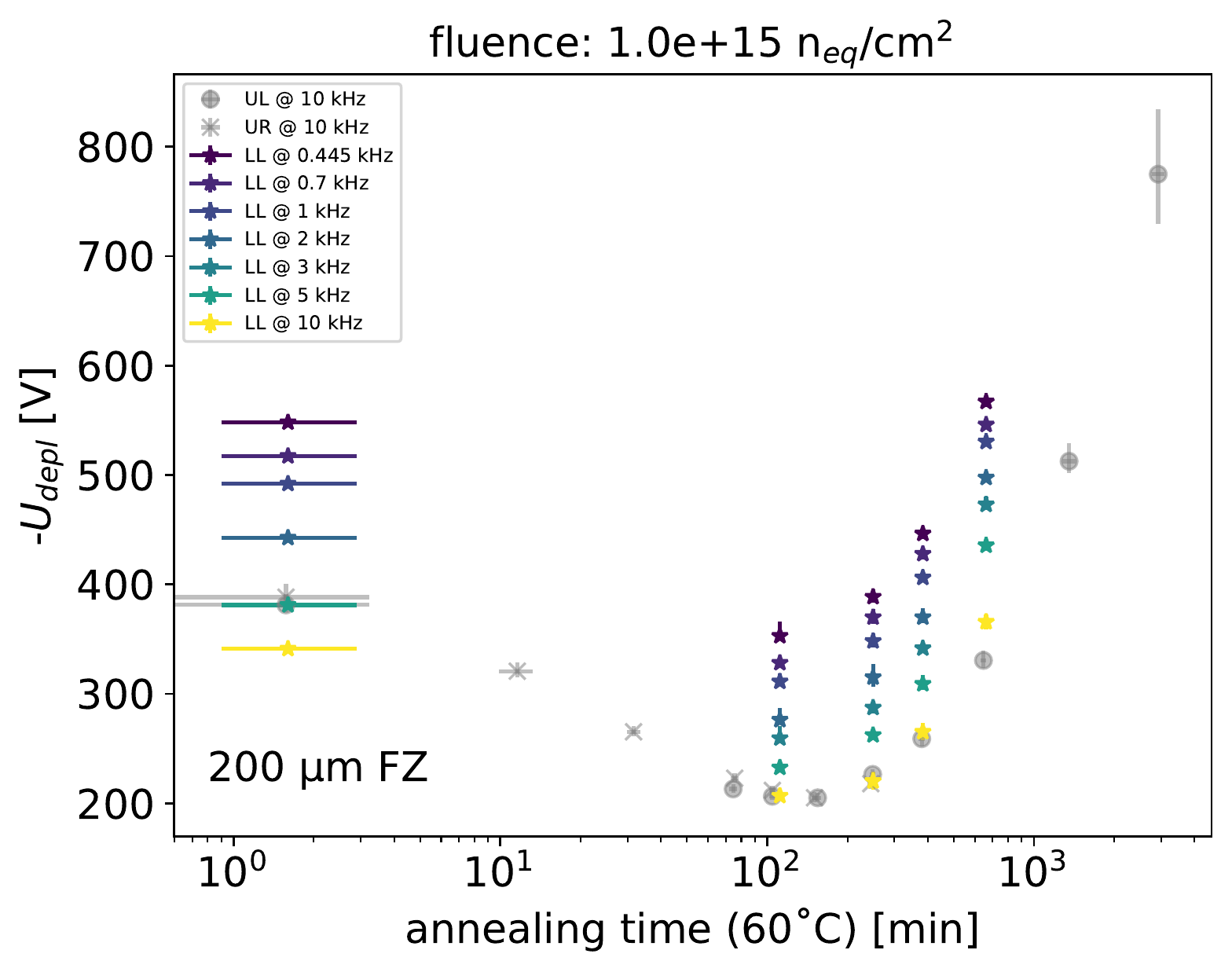}
    \caption{Depletion voltage for the 200\micron sample and a fluence of $1\cdot 10^{15} \neqcm$, measured at different frequencies.}
    \label{fig:fdep}
\end{figure}

\begin{figure}[htbp]
    \centering
    \includegraphics[width=0.8\textwidth]{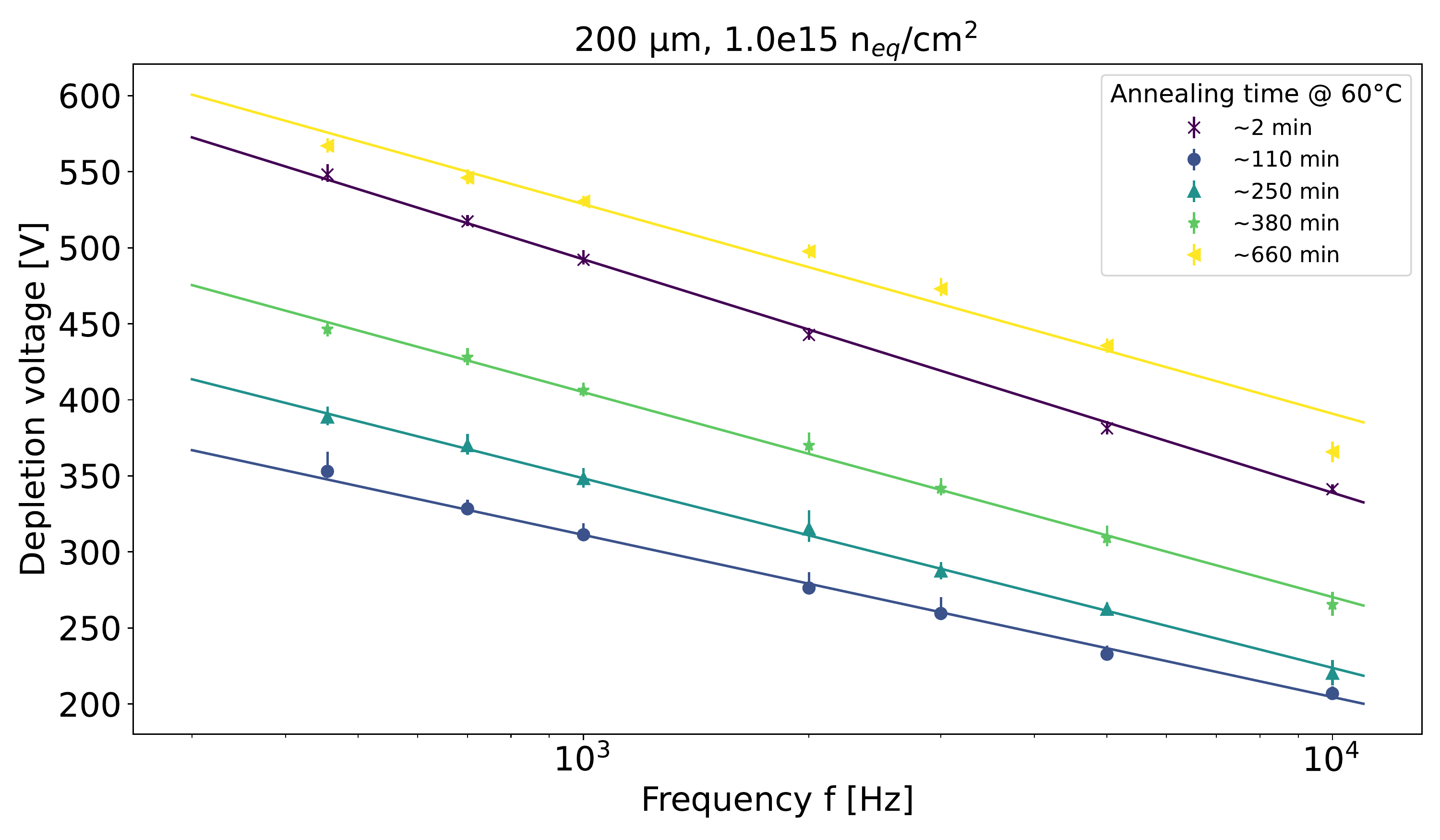}
    \caption{Frequency dependence of the depletion voltage for different annealing times for the sample 2002 LL.}
    \label{fig:fscan}
\end{figure}

\begin{figure}[htbp]
    \centering
    \includegraphics[width=0.8\textwidth]{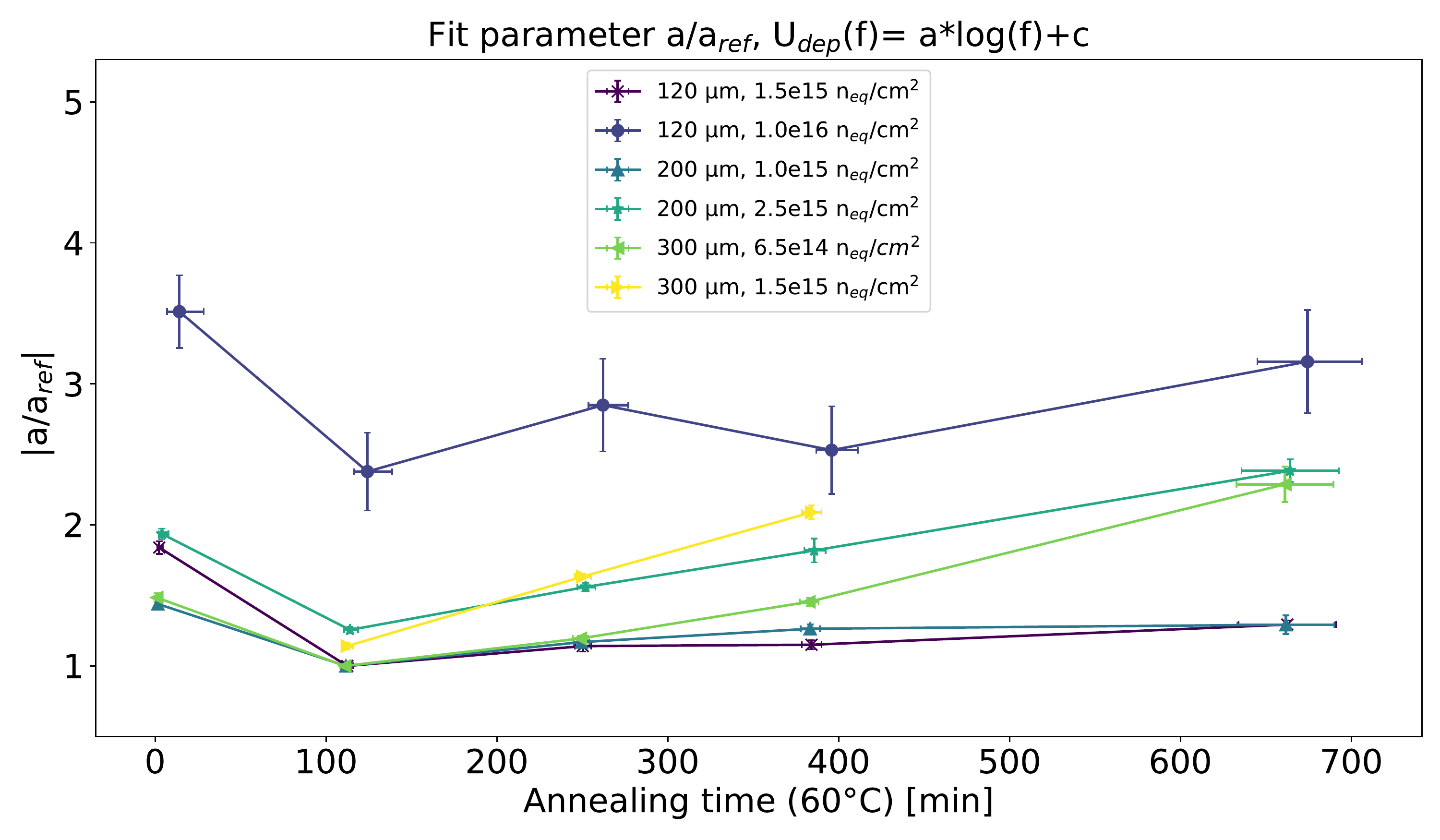}
    \caption{Frequency scaling factor a as a function of annealing time for different fluences and thicknesses, normalised to the scaling factor at 110\minutes of annealing and the lowest fluence sample for each thickness.}
    \label{fig:ascan}
\end{figure}

We select a subset of annealing points and the lowest and highest fluence samples per thickness for a dedicated study of this frequency dependence. These samples are listed in Table~\ref{tab:samples} as the ones originating from the lower left corner of the wafer. 
We consider five target annealing steps: 0, 110, 250, 380, and 660\minutes at 60\degreeC. The frequency is varied between 455\, Hz and 10\kiloHertz. As shown in Figure~\ref{fig:fdep}, the qualitative behaviour of the depletion voltage with annealing time persists for different frequencies. However, the depletion voltage decreases with the logarithm of the frequency, as illustrated in Figure~\ref{fig:fscan}, confirming the findings of Ref.~\cite{FdepCampbell}. 
As the data is sufficiently described by a function of the form 
\begin{equation}
    \mathrm{U_{dep} (f)} = a \cdot \log(\mathrm{f}) + c \text{,}
\end{equation}
where $f$ is the frequency and $a$ quantifies the dependence, the latter is evaluated for different annealing steps and fluences. The results are shown in Figure~\ref{fig:ascan}. 

The frequency scaling factor $a$ shows a significant dependence on the fluence as well as on the annealing time. Qualitatively, it follows the annealing behaviour of the depletion voltage, with larger factors at low and high annealing times, and with a minimum at about 100\minutes.
These results underline that there is no universal best frequency at a certain temperature for all fluences and annealing times contrary to what is suggested in Ref.~\cite{PETTERSON2007189}. Moreover, the results suggest that the Hamburg model parameters in general depend on the measurement frequency, also in terms of their fluence dependence. It is subject to further studies, also employing other methods to extract the depletion voltage than only from capacitance measurements, to investigate if in particular the fluence dependence of $g_a$ persists.

Another explanation for the qualitatively observed dependence could be a violation of the NIEL (Non Ionizing Energy Loss) scaling for this particular sensor material and fluence, which has previously been shown to be broken in other cases~\cite{Lindstrom:2002gb}, even though to our knowledge the material in this study is not as strongly enriched with oxygen or similar defects that could facilitate such a behaviour.
Second order effects in fluence are not modelled by the Hamburg model, which could lead to it not being precise enough at fluences beyond $10^{15}\neqcm$. Similar discrepancies have also been observed for the pixel detectors of the LHC experiments subjected to high fluences~\cite{Dawson:2764325,ATLAS:2021gld}, where the performance degradation is typically overestimated compared to the data. The high fluences could also produce a saturation of certain defects with increasing fluence and manifest as a decrease in the constants $g_a$ and $g_y$ and $g_c$. In particular the dominant contributions would be affected, in this case $g_a$ and $g_y$, as these second-order effects should scale with $N_{a/y}$ itself, they could become visible only for fluences beyond $10^{15} \neqcm$.



%
%
%
%

\section{Conclusions}
\label{seq:conclusions}

We present a comprehensive study of the capacitance and leakage current characteristics of silicion-diode test structures from 8" wafer material from the HGCAL prototype phase. The studied properties of the bulk material with respect to radiation damage and subsequent annealing are promising: the leakage current shows a behavior which is directly proportional to fluence, as expected, and decreasing consistently with time, and the depletion voltage exhibits only moderate reverse annealing, consistently for fluences ranging from $6.5\cdot 10^{14}$ to $10^{16}\neqcm$. 
For the float-zone material, the minimum depletion voltage is reached for annealing times of about 130\minutes at 60\degreeC, while the measurements suggest an optimal annealing time of about 80\minutes at 60\degreeC for the epitaxial material. Given the expected conditions the detector will be operated under, this means that annealing scenarios can be defined that ensure operation without entering a regime where the reverse annealing becomes dominant towards the end of the HL-LHC running. 


The current related damage rate \Ialpha is compatible between the different materials and with previous studies. 
The behaviour of the effective doping concentration, derived from the capacitance measurements, can be described by the Hamburg model parameterisation for each fluence individually. However, the measurements indicate signs of saturation for very high fluences or, in general, a break down of the model in the high fluence regime. 
Furthermore, we observe a significant fluence and annealing time dependence of the frequency scaling of the depletion voltage extracted from capacitance measurements.
These effects are  not fully understood yet and future studies that extend the studied annealing time and exploit less ambiguous methods for the extraction of the depletion voltage could shed light on them, such as e.g. the dependence of the charge collection on the bias voltage.

The results presented in this paper open the door for future studies necessary to derive a reliable parameterisation of the induced radiation damage and its annealing that is valid throughout the full HL-LHC running period. These studies could benefit from improved methods to extract the annealing parameters while relating the measurements at different fluences, annealing temperatures, and annealing times. Also, new parameterisations might be needed in general in this regime where the operating properties are determined not anymore predominantly by the depletion voltage, but more and more by charge trapping.

\begin{acknowledgments}

We thank the CERN EP-DT SSD group for providing their setups to perform the measurements described in this paper as well as for their input interpreting the measurements, in particular Michael Moll, Ruddy Costanzi, Esteban Curras Rivera, and Marcos Fernandez Garcia.
Furthermore, we thank the CMS HGCAL silicion group for useful discussions about the results presented here and the group at the Jo\v{z}ef Stefan Institute, Slovenia, for dedicated temperature measurements during irradiation. Moreover, we are grateful for feedback and discussions with Ronald Lipton, and suggestions from Nural Akchurin, Stathes Paganis, Rachel Yohay, Pedro Silva, Philippe Bloch, and Timo Peltola. This work has been sponsored by the Wolfgang Gentner Programme of the German Federal Ministry of Education and Research (grant no. 13E18CHA).

\end{acknowledgments}



\bibliographystyle{unsrt}  
\bibliography{main.bib}

\begin{thebibliography}{10}

\bibitem{Evans_2008}
Lyndon Evans and Philip Bryant.
\newblock {LHC} machine.
\newblock {\em Journal of Instrumentation}, 3(08):S08001--S08001, 2008.

\bibitem{Apollinari:2284929}
G.~Apollinari, I.~Béjar~Alonso, O.~Brüning, P.~Fessia, M.~Lamont, L.~Rossi,
  and L.~Tavian.
\newblock {\em {High-Luminosity Large Hadron Collider (HL-LHC): Technical
  Design Report V. 0.1}}.
\newblock CERN Yellow Reports: Monographs. CERN, Geneva, 2017.

\bibitem{CMSPFPaper}
{CMS Collaboration}.
\newblock Particle-flow reconstruction and global event description with the
  cms detector.
\newblock {\em Journal of Instrumentation}, 12(10):P10003–P10003, Oct 2017.

\bibitem{ATLASPF}
{ATLAS Collaboration}.
\newblock {Jet reconstruction and performance using particle flow with the
  ATLAS Detector}.
\newblock {\em Eur. Phys. J.}, C77(7), 2017.

\bibitem{Rovere:2020rqi}
Marco Rovere, Ziheng Chen, Antonio Di~Pilato, Felice Pantaleo, and Chris Seez.
\newblock {CLUE: A Fast Parallel Clustering Algorithm for High Granularity
  Calorimeters in High-Energy Physics}.
\newblock {\em Front. Big Data}, 3:591315, 2020.

\bibitem{qasim2022}
Shah~Rukh Qasim, Nadezda Chernyavskaya, Jan Kieseler, Kenneth Long, Oleksandr
  Viazlo, Maurizio Pierini, and Raheel Nawaz.
\newblock {End-to-end multi-particle reconstruction in high occupancy imaging
  calorimeters with graph neural networks}.
\newblock {\em Eur. Phys. J. C}, 82(8):753, 2022.

\bibitem{CMS_det_paper}
{CMS Collaboration}.
\newblock The {CMS} experiment at the {CERN} {LHC}.
\newblock {\em Journal of Instrumentation}, 3(08):S08004--S08004, aug 2008.

\bibitem{HGCAL-TDR}
{CMS Collaboration}.
\newblock {The Phase-2 Upgrade of the CMS Endcap Calorimeter}.
\newblock Technical Report CERN-LHCC-2017-023. CMS-TDR-019, 2017.

\bibitem{Peltola_2015}
T.~Peltola.
\newblock Silicon sensors for trackers at high-luminosity environment.
\newblock {\em Nuclear Instruments and Methods in Physics Research Section A:
  Accelerators, Spectrometers, Detectors and Associated Equipment}, 796:74--79,
  oct 2015.

\bibitem{CASSE2002465}
G~Casse, P.P Allport, T.J.V Bowcock, A~Greenall, M~Hanlon, and J.N Jackson.
\newblock First results on the charge collection properties of segmented
  detectors made with p-type bulk silicon.
\newblock {\em Nuclear Instruments and Methods in Physics Research Section A:
  Accelerators, Spectrometers, Detectors and Associated Equipment},
  487(3):465--470, 2002.

\bibitem{CASSE200646}
G.~Casse, P.P. Allport, and A.~Watson.
\newblock Effects of accelerated annealing on p-type silicon micro-strip
  detectors after very high doses of proton irradiation.
\newblock {\em Nuclear Instruments and Methods in Physics Research Section A:
  Accelerators, Spectrometers, Detectors and Associated Equipment},
  568(1):46--50, 2006.
\newblock New Developments in Radiation Detectors.

\bibitem{hamamatsu}
{Hamamatsu Photonics}.
\newblock \url{https://www.hamamatsu.com/eu/en.html}.

\bibitem{Lindstrom:1999mw}
G.~Lindstrom, M.~Moll, and E.~Fretwurst.
\newblock {Radiation hardness of silicon detectors: A challenge from
  high-energy physics}.
\newblock {\em Nucl. Instrum. Meth. A}, 426:1--15, 1999.

\bibitem{Moll:1999kv}
Michael Moll.
\newblock {\em {Radiation damage in silicon particle detectors: Microscopic
  defects and macroscopic properties}}.
\newblock PhD thesis, Hamburg U., 1999.

\bibitem{dimic_reactor_1978}
V.~Dimic.
\newblock {\em Reactor {TRIGA} at the {JStefan} institute in {Ljubljana}}.
\newblock Reaktor {TRIGA} {Instituta} ''{Jozef} {Stefan}'' v {Ljubljani}.
  Institut za Nuklearne Nauke Boris Kidric, Yugoslavia, 1978.

\bibitem{Cindro:2019cxd}
V.~Cindro et~al.
\newblock {Measurement of the charge collection in irradiated miniature sensors
  for the upgrade of the ATLAS phase-II strip tracker}.
\newblock {\em Nucl. Instrum. Meth. A}, 924:153--159, 2019.

\bibitem{SavGol}
A.~{Savitzky} and M.~J.~E. {Golay}.
\newblock {Smoothing and differentiation of data by simplified least squares
  procedures}.
\newblock {\em Analytical Chemistry}, 36:1627--1639, January 1964.

\bibitem{Boggs1989OrthogonalDR}
Paul~T. Boggs and Janet~R. Donaldson.
\newblock Orthogonal distance regression.
\newblock 1989.

\bibitem{Chilingarov:1511886}
A~Chilingarov.
\newblock {Generation current temperature scaling}.
\newblock Technical report, CERN, Geneva, 2013.

\bibitem{Moll:2640820}
Michael Moll.
\newblock {Displacement damage in silicon detectors for high energy physics.
  Displacement Damage in Silicon Detectors for High Energy Physics}.
\newblock {\em IEEE Trans. Nucl. Sci.}, 65(8):1561--1582, 2018.

\bibitem{Lindstrom:2002gb}
G.~Lindstrom.
\newblock {Radiation damage in silicon detectors}.
\newblock {\em Nucl. Instrum. Meth. A}, 512:30--43, 2003.

\bibitem{LDiehl_ptype}
Leena Diehl, Liv Wiik-Fuchs, Riccardo Mori, M.~Hauser, Karl Jakobs, and Ulrich
  Parzefall.
\newblock Annealing studies on irradiated p-type silicon strip sensors designed
  for the atlas phase ii tracking detector.
\newblock {\em RAD Association Journal}, 3, 01 2018.

\bibitem{FdepCampbell}
D.~Campbell, A.~Chilingarov, and Terry Sloan.
\newblock Frequency and temperature dependence of the depletion voltage from cv
  measurements for irradiated si detectors.
\newblock {\em Nuclear Instruments and Methods in Physics Research Section
  A-accelerators Spectrometers Detectors and Associated Equipment - NUCL
  INSTRUM METH PHYS RES A}, 492:402--410, 10 2002.

\bibitem{PETTERSON2007189}
M.K. Petterson, H.F.-W. Sadrozinski, C.~Betancourt, M.~Bruzzi, M.~Scaringella,
  C.~Tosi, A.~Macchiolo, N.~Manna, D.~Creanza, M.~Boscardin, C.~Piemonte,
  N.~Zorzi, L.~Borrello, A.~Messineo, and G.F. {Dalla Betta}.
\newblock Charge collection and capacitance–voltage analysis in irradiated
  n-type magnetic czochralski silicon detectors.
\newblock {\em Nuclear Instruments and Methods in Physics Research Section A:
  Accelerators, Spectrometers, Detectors and Associated Equipment},
  583(1):189--194, 2007.
\newblock Proceedings of the 6th International Conference on Radiation Effects
  on Semiconductor Materials, Detectors and Devices.

\bibitem{Dawson:2764325}
I~Dawson.
\newblock {\em {Radiation effects in the LHC experiments: Impact on detector
  performance and operation}}.
\newblock CERN Yellow Reports: Monographs. CERN, Geneva, 2021.

\bibitem{ATLAS:2021gld}
Georges Aad et~al.
\newblock {Measurements of sensor radiation damage in the ATLAS inner detector
  using leakage currents}.
\newblock {\em JINST}, 16:P08025, 2021.

\end{thebibliography}

\end{document}